\documentclass[twocolumn,showpacs,preprintnumbers,amsmath,amssymb,superscriptaddress,pra,longbibliography]{revtex4-1}

\usepackage{graphicx}
\usepackage{dcolumn}
\usepackage{bm}
\usepackage{graphics}
\usepackage{subfigure}
\usepackage{graphicx}
\usepackage{epsfig}
\usepackage{epstopdf}
\usepackage{xcolor}

\allowdisplaybreaks

\begin{document}

\title{High-temperature expansion of the viscosity in interacting quantum gases}

\author{Johannes Hofmann}
\email{jbh38@cam.ac.uk}
\affiliation{Department of Applied Mathematics and Theoretical Physics, Centre for Mathematical Sciences, University of Cambridge,  Cambridge CB3 0WA, United Kingdom}
\affiliation{TCM Group, Cavendish Laboratory, University of Cambridge, Cambridge CB3 0HE, United Kingdom}

\date{\today}

\begin{abstract}
We compute the frequency-dependent shear and bulk viscosity spectral functions of an interacting Fermi gas in a quantum virial expansion up to second quadratic order in the fugacity parameter $z=e^{\beta \mu}$, which is small at high temperatures. Calculations are carried out using a diagrammatic finite-temperature field-theoretic framework, in which the analytic continuation from Matsubara to real frequencies is carried out in closed analytic form. Besides a possible zero-frequency Drude peak, our results for the spectral functions show a broad continuous spectrum at all frequencies with an additional bound-state contribution for frequencies larger than the dimer-breaking energy. Our results are consistent with various sum rules and universal high-frequency tails. In the low-frequency limit, the shear viscosity spectral function is recast as a collision integral, which reproduces known results for the static shear viscosity from kinetic theory. Our findings for the static bulk viscosity of a Fermi gas near unitarity, however, show a nonanalytic dependence on the scattering length, at variance with kinetic theory.
\end{abstract}

\pacs{}
\maketitle

\section{Introduction}

Strongly interacting quantum gases form a paradigmatic quantum many-body system in which the interaction strength and trap parameters are tuned at will~\cite{bloch08,giorgini08}. Properties of cold gases are probed accurately by studying their collective oscillations and their expansion dynamics. The damping or dissipation in these processes is set by the shear viscosity $\eta$ and bulk viscosity $\zeta$~\cite{schaefer12,schaefer14}, the experimental determination of which has been an active topic over the past decade~\cite{cao11,cao11b,vogt12,elliott14,joseph15}. Indeed, strongly interacting Fermi gases are said to be ``perfect fluids'' in which both bulk and shear viscosity are anomalously small~\cite{chafin13, schaefer14}. The shear viscosity comes close to a conjectured lower limit for the shear viscosity to entropy ratio $\eta/s \sim \hbar/k_B$~\cite{kovtun05}, which should apply to strongly interacting quantum systems close to a scale-invariant point. Likewise, the bulk viscosity will vanish in scale-invariant systems~\cite{son07}.

Formally, the shear and bulk viscosities are given as the zero-frequency limit of corresponding viscosity spectral functions $\eta(\omega)$ and $\zeta(\omega)$, which are defined in terms of Kubo relations for the retarded stress tensor correlation function
\begin{align}
\eta(\omega) &= \frac{{\rm Im} \, \chi_{xy,xy}^R(\omega)}{\omega} \label{eq:defshear}, \\
\zeta(\omega) &= \frac{{\rm Im} \, \chi_{ii,jj}^R(\omega)}{d^2 \omega} \label{eq:defbulk} ,
\end{align}
where $d$ is the dimension of the gas. The retarded response function $\chi_{ij,kl}^R(\omega)$ is the Fourier transform of the real-time response $\chi_{ij,kl}^R(t) = - i \hbar^{-1} \Theta(t) \langle [\hat{\Pi}_{ij}(t), \hat{\Pi}_{kl}(0)]\rangle$, with $\langle \cdots \rangle$ the thermal average and $\hat{\Pi}_{ij}$ the stress tensor. In Eq.~\eqref{eq:defbulk} there is a summation convention for  the doubly occurring space indices $i$ and $j$. Quite generally, the extrapolation of numerical results for the response functions to zero frequency is an ill-posed problem that requires numerical extrapolation methods. A first-principles calculation of the shear and bulk viscosity from Kubo relations is thus intricate~\cite{enss11,wlazlowski12,wlazlowski13,wlazlowski15}. The frequency-dependent viscosity correlation functions are not only important in theoretical studies to obtain the static limit, however. Linear combinations are also related to the long-wavelength limit of the dynamic structure factor~\cite{taylor10}, which can be measured using Bragg spectroscopy~\cite{veeravalli08,carcy19}. Moreover, as discussed further in this paper, the bulk viscosity spectral function describes the energy absorption in response to an oscillating scattering length~\cite{son10}.

Exact constraints on the spectral functions~\eqref{eq:defshear} and~\eqref{eq:defbulk} exist in the form of sum rules and universal high-frequency tails~\cite{taylor10,hofmann11,goldberger12}. At large frequencies, the spectral functions have universal power-law high-frequency tails that decay as $\eta(\omega), \zeta(\omega) \stackrel{\omega\to\infty}{\longrightarrow} C \omega^{-\frac{4-d}{2}}$~\cite{taylor10,hofmann11,goldberger12}, with a magnitude set by the contact parameter $C$~\cite{tan08a,tan08b,tan08c}. Moreover, the total integrated spectral weight depends on the derivative of the contact with respect to the scattering length~\cite {taylor10,taylor12}. These exact results should be obeyed by any calculation of spectral functions.

An often-used framework to compute transport coefficients in quantum gases is kinetic theory~\cite{bruun05,massignan05,bruun07,schaefer12b,enss12,bruun12,dusling13}. However, strictly speaking, kinetic theory only applies in some limiting cases, such as the low-temperature limit of a strongly spin-imbalanced Fermi gas, which is described by Fermi liquid theory~\cite{pines66,abrikosov75,recati12}. Here a kinetic description exists for long-lived quasiparticle excitations derived from an underlying density functional~\cite{recati10,hofmann18}. For the spin-balanced gas, kinetic theory applies at both very low~\cite{landau49} and very high temperatures~\cite{yan19}. In this paper we study the high-temperature or nondegenerate limit, in which the thermal wavelength $\lambda_T=\sqrt{2\pi\hbar^2\beta/m}$ is much smaller than the interparticle distance, $n\lambda_T^d \ll 1$, where $n$ is the density, $\beta$ inverse temperature, and $m$ the mass of an atom. A drawback of kinetic theory is that the high-frequency tails of the viscosity spectral functions are not correctly captured and the extrapolation to smaller temperatures is not controlled. While a formal correspondence between a kinetic description and the classical limit of the microscopic theory  is established within the Keldysh formalism~\cite{kamenev13}, a direct link between the results for transport coefficients obtained within kinetic theory and microscopic calculations is not apparent.

In this paper we consider the high-temperature limit of the viscosity spectral functions in a microscopic calculation using a quantum virial expansion. The virial expansion applies in the grand canonical ensemble and expands the thermal expectation value of an operator $\hat{\cal A}$ as a sum over expectation values restricted to the $N$-particle sector
\begin{align}
\langle \hat{\cal A} \rangle &= \frac{1}{\cal Z} {\rm Tr} [ e^{-\beta (\hat{H} - \mu N)} \hat{\cal A}  ] = \sum_N z^N {\rm Tr}_{N,c} [e^{-\beta \hat{H}} \hat{\cal A} ] , \label{eq:virial}
\end{align}
where the expansion parameter $z=e^{\beta \mu}$ is called the fugacity. Here ${\rm Tr}_{N,c}$ denotes the connected trace restricted to the $N$-particle sector, i.e., 
\begin{align}
{\rm Tr}_{1,c} [e^{-\beta \hat{H}} \hat{\cal A} ] &= {\rm Tr}_{1} [e^{-\beta \hat{H}} \hat{\cal A} ] \\
{\rm Tr}_{2,c} [e^{-\beta \hat{H}} \hat{\cal A} ] &= {\rm Tr}_{2} [e^{-\beta \hat{H}} \hat{\cal A} ] - {\rm Tr}_{1} [e^{-\beta \hat{H}} \hat{\cal A} ] {\rm Tr}_{1} [e^{-\beta \hat{H}}] ,
\end{align}
and so on, where ${\rm Tr}_{N}$ sums all $N$-particle states. Since the $N$th-order term of Eq.~\eqref{eq:virial} involves only matrix elements of $N$-particle states, the virial expansion provides a link between the high-temperature properties of the quantum gas and few-particle solutions, which are often known. In particular, for the particle density, we obtain
\begin{align}
n \lambda_T^d &= z + 2 b_2 z^2 + 3 b_3 z^3 + \cdots  , \label{eq:density}
\end{align}
with $b_i$ the virial coefficients~\cite{liu13}. Truncating the expansion~\eqref{eq:virial} or~\eqref{eq:density} after the first few terms holds for $z\ll1$, which corresponds to the nondegenerate or high-temperature regime $n \lambda_T^d\ll1$. Note that the quantum virial expansion is valid for any interaction strength, and in this sense it is a nonperturbative method. In order to make contact with many-body theory, in this paper we use a diagrammatic method to compute the viscosity spectral functions.

\begin{figure}[t]
\scalebox{0.9}{\includegraphics{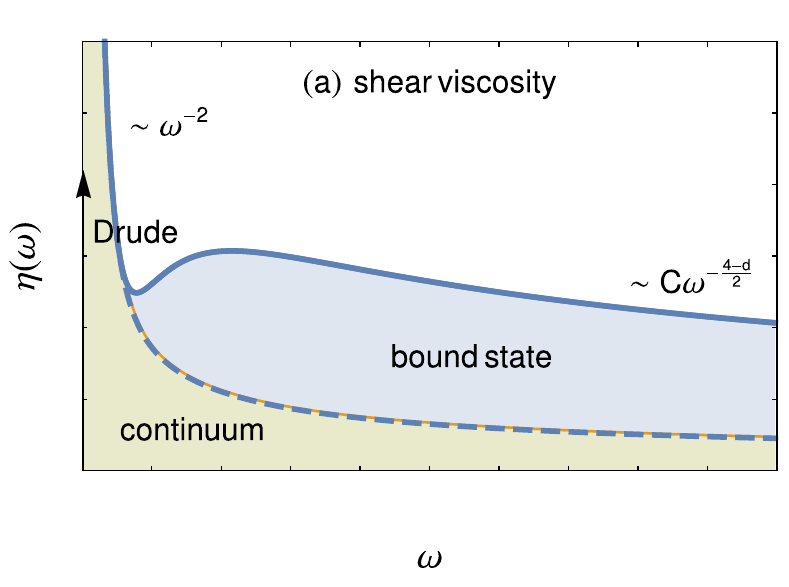}}
\scalebox{0.9}{\includegraphics{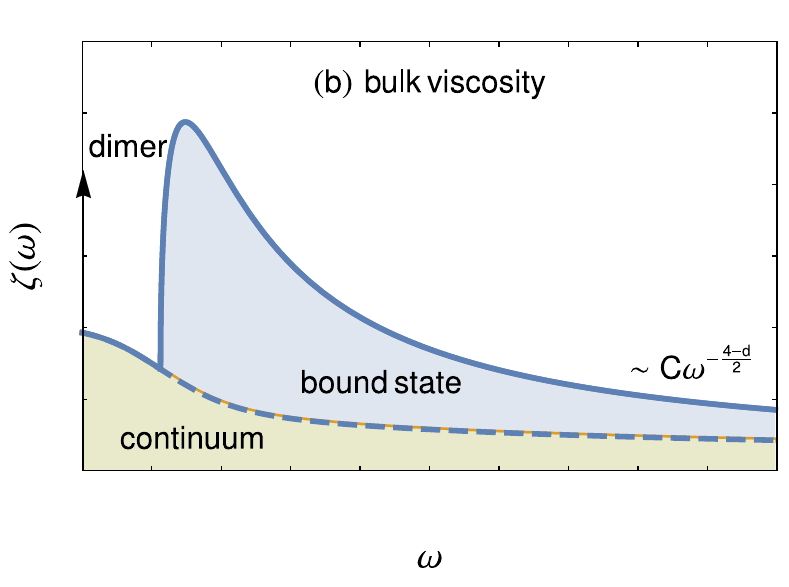}}
\caption{Structure of the (a) shear viscosity and (b) bulk viscosity spectral function. There is a continuum contribution for all frequencies, which diverges as a power law at low frequencies for the shear viscosity spectral function and saturates to a constant value for the bulk viscosity. At zero frequency, there is a Drude $\delta$ contribution indicated by the black arrow. When present, there is a bound-state contribution above a threshold frequency. At large frequencies, both continuum and bound-state parts contribute to a universal power-law tail.}
\label{fig:1}
\end{figure}

The main result of this paper is the exact viscosity spectral functions $\eta(\omega)$ and $\zeta(\omega)$ up to second order in the fugacity. An advantage of our calculation is that the analytical continuation to real frequencies is performed exactly and does not require extrapolation schemes. For illustration, Fig.~\ref{fig:1} shows the general form of the spectral functions obtained in this paper. There is a broad continuous spectrum at all frequencies, which arises from interactions between scattering atoms. If two-particle bound states are present, there is additional weight from bound-free transitions, which contribute at frequencies larger than the bound-state energy with a nonanalytic frequency dependence right above threshold. In particular, for the bulk viscosity, the bound-state part has a very steep onset and dominates the spectral function. At small nonzero frequencies, only the continuum part remains. For the shear viscosity, this continuum part diverges as a power law $\lim_{\omega \to 0} \eta(\omega) \sim 1/\omega^2$, whereas for the bulk viscosity, it saturates to a finite value. There is an additional $\delta$-peak contribution at zero frequency for both shear and bulk viscosity. For the bulk viscosity, this peak arises from bound-bound transitions and has nonzero weight only if a bound state is present. Both the continuum and bound-state parts contribute at high frequencies. 

It turns out that the interacting contribution to the low-frequency divergence of the shear viscosity spectral function takes the form of a collision integral, which allows us to obtain a static shear viscosity using a memory function resummation. Our results agree with calculations of the shear viscosity within kinetic theory. Interestingly, in the absence of bound states in the two-particle sector, the static limit of the bulk viscosity spectral function can be taken directly. We quote a result of our calculation in 3D near the unitary limit
\begin{align}
\lim_{a \to -\infty} \frac{\lambda_T^3}{\hbar z^2} \zeta |_{\omega\to 0} &= \dfrac{\sqrt{2} }{9 \pi^2} \biggl(\frac{\lambda_T}{a}\biggr)^2 \Bigl(\ln \dfrac{2 \pi e^{-\gamma_E} a^2}{\lambda_T^2} - 1\Bigr) , \label{eq:bulkunitary}
\end{align}
where the zero-frequency limit of the spectral function was taken before the unitary limit, with $a$ the scattering length and $\gamma_E$ the Euler-Mascheroni constant. This result differs from kinetic theory, which predicts a quadratic dependence~$\lambda_T^3 \zeta/\hbar z^2 \sim (\lambda_T/a)^2$~\cite{dusling13}, but it is consistent with a predicted critical scaling near unitarity~\cite{maki18,maki19a}.

This paper is structured as follows: At the beginning of Sec.~\ref{sec:II} we introduce the zero-range model and discuss the Kubo formula that define the spectral functions. The remainder of Sec.~\ref{sec:II} then presents the main calculation of this paper. In Sec.~\ref{sec:IIa} we set up a diagrammatic framework to compute the high-temperature expansion of correlation functions exactly to second order in the fugacity $z$. The diagrammatic calculation for the stress correlator and the contact correlator used to compute the viscosity spectral functions is presented in Secs.~\ref{sec:IIb} and~\ref{sec:IIc}, with details of the analytical continuation from Matsubara space to real frequencies relegated to Appendix~\ref{app:1}. The results for the shear and bulk viscosity spectral functions are presented in Sec.~\ref{sec:III}. We discuss the general form of the shear viscosity spectral function in Sec.~\ref{sec:IIIa}, with particular attention to the universal high-frequency behavior. We show how a memory function approach reproduces the kinetic results in the static limit. The bulk viscosity spectral function is discussed in Sec.~\ref{sec:IIIb}. Furthermore, we demonstrate that our results are consistent with universal high-frequency tails and sum rule constraints. In addition, we provide analytic results for the static bulk viscosity and comment on the discrepancy with kinetic theory. The paper is concluded by a summary and outlook in Sec.~\ref{sec:IV}.

\section{High-temperature expansion of the viscosity spectral function}\label{sec:II}

We consider a Fermi quantum gas with zero-range interactions described by the Hamiltonian
\begin{align}
\hat{H} &= \int d^dx \, \biggl\{\sum_\sigma \hat{\psi}_\sigma^\dagger \frac{-\hbar^2\nabla^2}{2m} \hat{\psi}_\sigma^{} - g \hat{\psi}_\uparrow^\dagger \hat{\psi}_\downarrow^\dagger \hat{\psi}_\downarrow^{} \hat{\psi}_\uparrow^{} \biggr\} , \label{eq:zerorange}
\end{align}
where $\sigma = \uparrow,\downarrow$ and $g$ is the bare interaction strength. The stress tensor $\hat{\Pi}_{ij}$ in the zero-range model~\eqref{eq:zerorange} takes the form (the summation over $\sigma$ is implied)
\begin{align}
\hat{\Pi}_{ij} &= \frac{\hbar^2}{2m} \bigl[(\nabla_i \hat{\psi}_\sigma^\dagger) (\nabla_j \hat{\psi}_\sigma) + (\nabla_j \hat{\psi}_\sigma^\dagger) (\nabla_i \hat{\psi}_\sigma)\bigr] \nonumber \\
&\quad - \delta_{ij} \Bigl[\frac{\hbar^2}{4m} \nabla^2 (\hat{\psi}_\sigma^\dagger \hat{\psi}_\sigma) + g \hat{\psi}_\uparrow^\dagger \hat{\psi}_\downarrow^\dagger \hat{\psi}_\downarrow^{} \hat{\psi}_\uparrow^{}\Bigr] . \label{eq:stresstensor}
\end{align}
This expression is derived, for example, from the Euler equation in operator form, i.e., the Heisenberg equation of motion for the current operator, $- \frac{i m}{\hbar} [\hat{j}_i, \hat{H}] = - \partial_j \hat{\Pi}_{ij}$, where $\hat{j}_i = \frac{-i\hbar}{2m} [\psi^\dagger_\sigma (\partial_i \psi) - (\partial_i \psi^\dagger_\sigma)  \psi]$ and the Hamiltonian is given in Eq.~\eqref{eq:zerorange}. The off-diagonal terms of the stress tensor as used in the shear viscosity spectral function~\eqref{eq:defshear} only involve bilinear operators. For direct calculations of the bulk viscosity, however, the expression~\eqref{eq:defbulk} is inconvenient as the stress tensor trace involves both bilinear and quartic operators:
\begin{align}
\hat{\Pi}_{ii} &= \frac{\hbar^2}{m}
 \Bigl[(\nabla_i \hat{\psi}_\sigma^\dagger) (\nabla_i \hat{\psi}_\sigma) - \frac{d}{4} \nabla^2 (\hat{\psi}_\sigma^\dagger \hat{\psi}_\sigma)\Bigr] 
 - d g \hat{\psi}_\uparrow^\dagger \hat{\psi}_\downarrow^\dagger \hat{\psi}_\downarrow^{} \hat{\psi}_\uparrow^{} .
\end{align}
We use a summation convention where the sum over the space index $i$ is implied. It was established in Refs.~\cite{martinez17,czajka17,fujii18} that the bulk viscosity is also defined in terms of the response function
\begin{align}
\zeta(\omega) &= \frac{{\rm Im} \, \chi_{{\cal O},{\cal O}}^R(\omega)}{\omega} , \label{eq:defbulkbetter} 
\end{align}
with ${\cal O} = \frac{1}{d} (\hat{\Pi}_{ii} - 2 \hat{\cal E})$, where $\hat{\cal E}$ is the energy density with $\hat{H} = \int d^dx \, \hat{\cal E}$. The operator $\hat{\cal O}$ is linked to the Tan contact operator and is a measure of the breaking of scale invariance in the quantum gas~\cite{hofmann12}
\begin{align}
\hat{\cal O} &= \begin{cases}
\dfrac{a \hbar^2 \hat{C}}{2 m} & {\rm (1D)} \\[1.5ex]
\dfrac{\hbar^2 \hat{C}}{4 \pi m} & {\rm (2D)} \\[1.5ex]
\dfrac{\hbar^2 \hat{C}}{12 \pi m a} & {\rm (3D)} ,
\end{cases} \label{eq:defO}
\end{align}
where $a$ is the effective scattering length in $d$ dimensions and ${\hat C}$ is the contact operator~\cite{tan08a,tan08b,tan08c}, ${\hat C} = \frac{m^2 g^2}{\hbar^4} \hat{\psi}_\uparrow^\dagger \hat{\psi}_\downarrow^\dagger \hat{\psi}_\downarrow^{} \hat{\psi}_\uparrow^{}$ in the zero-range model~\cite{braaten08a,braaten08b}. The contact parametrizes universal short-range correlations and thermodynamic properties and is thus a central experimental observable~\cite{sagi12,hoinka13,carcy19,mukherjee19}. The expectation value of the contact sets the magnitude of the leading short-distance divergence of the pair correlation function; hence intuitively it describes the number of pairs with opposite spin in close proximity~\cite{braaten12}. The contact is linked to the partition function of the gas by the adiabatic relation~\cite{tan08c,barth11,valiente11,braaten12}
\begin{align}
C &= \begin{cases}
\dfrac{2 m}{\hbar^2} \dfrac{\partial (\Omega/V)}{\partial a} & {\rm (1D)} \\[1.5ex]
\dfrac{2 \pi m}{\hbar^2} a \dfrac{\partial (\Omega/V)}{\partial a} & {\rm (2D)} \\[1.5ex]
\dfrac{4 \pi m}{\hbar^2} a^2 \dfrac{\partial (\Omega/V)}{\partial a} & {\rm (3D)} .
\end{cases}
\end{align}
Indeed, the contact also governs the nonanalytic short-time correlation of the stress correlator $\chi_{ij,kl}^R(t)$, and thus sets the magnitude of a power-law high-frequency tail of the viscosity spectral functions~\cite{taylor10,hofmann11,goldberger12}, which will be discussed later in this paper. 

The expectation value of ${\hat O}$ sets the deviation of the pressure ${\cal P}$ from the scale-invariant value: ${\cal P} - \frac{2}{d} {\cal E} = \langle \hat{O} \rangle$. In two dimensions, the operator $\hat{\cal O}$ indicates a quantum scale anomaly, where formally it describes an anomalous contribution that opens up the nonrelativistic conformal operator algebra of a scale-invariant quantum gas~\cite{hofmann12}. Indeed, the bulk viscosity spectral function vanishes identically in a scale-invariant system and is nonzero only when scale invariance is broken~\cite{son07}. On a technical level, the contact response $\chi_{{\cal O},{\cal O}}^R$ does not contain bilinear operators, which simplifies calculations of the bulk viscosity considerably. Note that the viscosity spectral functions can also be defined in terms of the transverse and longitudinal current response functions~\cite{forster90, taylor10, hofmann11}
\begin{align}
\eta(\omega) = \lim_{q\to 0} \frac{m\omega^2}{q^2} {\rm Im} \chi_T(\omega, q) , \\
\zeta(\omega) + \frac{2 (d-1)}{d} \eta(\omega) = \lim_{q\to 0} \frac{m\omega^2}{q^2} {\rm Im} \chi_L(\omega, q) ,
\end{align}
where $\chi_L$ and $\chi_T$ are the longitudinal and transverse parts of the current response function, $\chi^{JJ}_{ij} = \chi_L \frac{q_i q_j}{q^2} + \chi_T (\delta_{ij} - \frac{q_i q_j}{q^2})$. 
However, this definition requires the computation of correlation functions not only at finite external frequency but also nonzero momentum, which complicates the calculation in the present case.

Both the shear viscosity spectral function and the bulk viscosity spectral function can be measured in experiments. Since the longitudinal current response function is related to the density response function (and thus the dynamic structure factor) by the continuity equation, the linear combination $\zeta(\omega) + \frac{2 (d-1)}{d} \eta(\omega)$ is measured in Bragg spectroscopy experiments~\cite{veeravalli08,carcy19}. In particular, in scale-invariant systems (such as the unitary Fermi gas), where the bulk viscosity vanishes, the dynamic structure factor is proportional to the shear viscosity. Moreover, the contact correlation function used in Eq.~\eqref{eq:defbulkbetter} to define the bulk viscosity determines the response to an oscillating modulation of the inverse scattering length, i.e., a perturbation of the form $\delta \alpha(t) = \delta a^{-1}(t) = \alpha_0 \cos \omega t$~\cite{son10}. Such a perturbation changes the Hamiltonian as
\begin{align}
\delta H &= (a d) \delta \alpha(t) \hat{\cal O} .
\end{align}
The rate at which energy is absorbed by the system in response to this perturbation, for example, is proportional to the contact correlation function
\begin{align}
\frac{dE}{dt} &= \frac{a^2 d^2}{2} \alpha_0^2 \, \omega \, {\rm Im} \, \chi_{{\cal O},{\cal O}}^R(\omega) .
\end{align}

\subsection{Diagrammatic virial expansion}\label{sec:IIa}

Within finite-temperature field theory, we compute the discrete Fourier transform of the time-ordered correlation function
\begin{align}
\chi_{\cal A,B}(i\omega_n) &= \int_0^\beta d\tau \, e^{i \hbar \omega_n \tau} \langle {\cal T}_\tau {\cal A}(0) {\cal B}(\tau) \rangle ,
\end{align}
where $i \omega_n$ is a bosonic Matsubara frequency and ${\cal T}_\tau$ denotes time ordering with operators at later times placed to the left. Retarded response functions as needed for the spectral functions~\eqref{eq:defshear} and~\eqref{eq:defbulkbetter} are obtained by analytically continuing from Matsubara to real frequencies $i\omega_n\to\omega+i0$~\cite{mahan00}. When working in Matsubara space, the fugacity parameter $z=e^{\beta \mu}$ is contained in Bose and Fermi factors used to convert Matsubara frequency summations to contour integrals~\cite{mahan00}. An expansion in $z$ is not apparent in this framework as it changes the analytic structure of the integrand and hence does not commute with the frequency integration. Little is gained by computing the full result first and then expanding in the fugacity.

It turns out that a fugacity expansion can be developed systematically when working in imaginary time as opposed to Matsubara space. There the fugacity only enters through the imaginary-time single-particle Green's function
\begin{align}
G(\tau, {\bf k}) &= - e^{- \tau (\varepsilon_{\bf k} - \mu)} \bigl[\Theta(\tau) - n_F(\varepsilon_{\bf k} - \mu) \bigr] , \label{eq:greens}
\end{align}
where ${\bf k}$ is the wave vector, $\varepsilon_{\bf k} = \frac{\hbar^2k^2}{2m}$ the single-particle energy, and $n_F(\varepsilon_{\bf k} - \mu)$ the Fermi factor, which can be expanded directly as $n_F(\varepsilon_{\bf k} - \mu) = -\sum_{n=1}^\infty (-1)^n z^n e^{- n \beta \varepsilon_{\bf k}}$. This expansion commutes with any internal integration over imaginary time or momenta. Diagrammatic calculations of the equation of state date back to the early days of quantum many-body theory with works by Montroll and Ward~\cite{montroll58} and Vedenov and Larkin~\cite{vedenov58} for the screened plasma. For quantum gases with short-range interactions, a diagrammatic framework was put forward in Refs.~\cite{kaplan11,leyronas11}, which compute the third virial coefficient of a strongly interacting Fermi gas. These techniques are not restricted to thermodynamic quantities but can also be applied to correlation functions as well. Indeed, besides virial coefficients~\cite{kaplan11,leyronas11,ngampruetikorn15,barth15}, diagrammatic calculations have been performed for the spectral function~\cite{barth14,ngampruetikorn13,sun15,sun17}, the momentum distribution of Bose gases including three-body Efimov effects~\cite{barth15}, and also the electron gas~\cite{hofmann13}. Most of these works employ a framework similar to the one introduced in Ref.~\cite{leyronas11}. In this section we collect the Feynman rules as used in the remainder of this paper.

\begin{figure}[t!]
\begin{center}
\raisebox{-0.44cm}{\scalebox{0.6}{\includegraphics{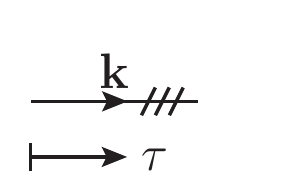}}} \hspace{-0.4cm}$= G^{(m)}(\tau, {\bf k})$ \qquad
\raisebox{-0.4cm}{\scalebox{0.6}{\includegraphics{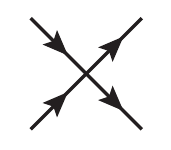}}}  $= - g$ \\
\subfigure[]{\raisebox{-0.06cm}{\scalebox{0.4}{\includegraphics{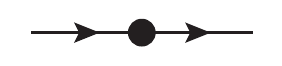}}} \hspace{0.cm}$= T^0_{xy}({\bf k}) = \tfrac{\hbar^2}{m} k_x k_y$ \qquad
\raisebox{-0.4cm}{\scalebox{0.6}{\includegraphics{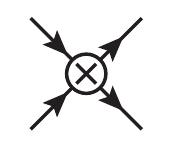}}}  $= \tfrac{m^2 g^2}{\hbar^4}$ 
\label{fig:2a}}\\
\subfigure[]{\scalebox{0.6}{\includegraphics{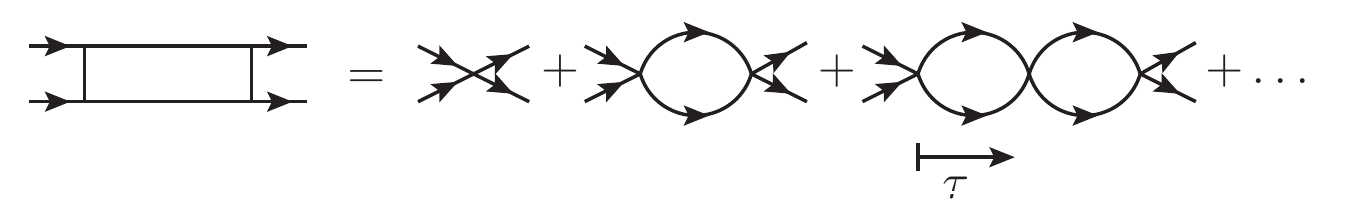}}\label{fig:2b}}
\caption{(a) Feynman rules of the zero-range model in imaginary time. Imaginary time runs from the left-hand side to the right-hand side. (b) Ladder diagrams that describe two-particle scattering to leading order ${\cal O}(z^0)$. The sum of all repeated scattering processes gives the vacuum $T$ matrix.}
\label{fig:2}
\end{center}
\end{figure}

The Feynman rules for the model~\eqref{eq:zerorange} in imaginary time are as follows. Imaginary time runs from the left to the right of the Feynman diagram in an interval $[0,\beta)$. Every vertex is assigned an imaginary-time index $\tau$ and contributes a factor $- g$. In addition, we denote the bare off-diagonal stress tensor insertion~\eqref{eq:stresstensor} on a propagator at a specific imaginary time by
\begin{align}
T^0_{xy}({\bf k}) = \frac{\hbar^2}{m} k_x k_y .
\end{align}
Likewise, the contact operator insertion carries a factor of $\frac{m^2 g^2}{\hbar^4}$. As shown in Fig.~\ref{fig:2a}, we indicate the ${\cal O}(z^m)$ term of the single-particle Green's function~\eqref{eq:greens} by a line that is slashed $m$ times. It contributes factors of
\begin{align}
G^{(m)}(\tau, {\bf k}) &= \begin{cases}
 - e^{\tau \mu} \, e^{- \tau \varepsilon_{\bf k}} \, \Theta(\tau) & m=0 \\[1.5ex]
 (-1)^{m-1} z^m e^{\tau \mu} \, e^{- (m \beta + \tau) \varepsilon_{\bf k}} & m\geq 1 .
 \end{cases} \label{eq:feynmanprop}
\end{align}
Contributions ${\cal O}(z^N)$ to some correlation function are represented by Feynman diagrams that contain $N$ slashes. 
A key observation is that the leading order in $z$ of the propagator is purely retarded, i.e., it runs from the left to right in any Feynman diagram.  Leading-order diagrams thus contain a minimum number of backward-propagating (advanced) propagators. We impose momentum conservation and integrate over internal momenta and imaginary-time labels. Imaginary-time integrals take the form of convolution integrals which are simplified using the convolution theorem for the Laplace transform
\begin{align}
&\int_0^\infty d(t_1,\ldots,t_n) \nonumber \\
&\quad \times [\Theta(t_1) f_1(t_1)] \cdots [\Theta(t_n) f_n(t_n)] \Theta(\tau - t_1 - \ldots - t_n) \nonumber \\
&= \int_{\rm Bw} \frac{ds}{2\pi i} e^{- \tau s} f_1(s) \cdots f_n(s) , \label{eq:convolution}
\end{align}
where we define the Laplace transform $f(s) = \int_0^\infty dt \, e^{st} f(t)$ with inverse  $f(t) = \int_{\rm Bw} \frac{ds}{2\pi i} \, e^{-st} f(t)$. Here Bw is the Bromwich contour which runs from negative to positive infinity with all singularities of the integrand to the right of the contour.

Typical Feynman diagrams constructed as described above involve repeated few-body scattering processes as subblocks. It turns out that to leading order in the fugacity ${\cal O}(z^0)$, the sum of these processes gives the free-particle scattering matrices. These blocks link the high-temperature properties of an interacting many-body system to known few-body solutions. We will use the  two-particle scattering matrix $T_2$, which we represent by a box as shown in Fig.~\ref{fig:2b}. Using the convolution theorem and performing the Laplace transform, $T_2$ is given by a geometric series~\cite{barth15b}
\begin{align}
T_2^{-1}(s, {\bf P}) &= - \frac{1}{g} - L_{\bf P}(s) , \label{eq:defT}
\end{align}
where ${\bf P}$ is the center-of-mass momentum and we define the bare two-particle propagator
\begin{align}
L_{\bf P}(\tau) &= \Theta(\tau) \, \int \frac{d^dl}{(2\pi)^d} \, e^{- \tau (\varepsilon_{\bf l} + \varepsilon_{\bf P-l})} ,\label{eq:defL1} \\
L_{\bf P}(s) &= \int \frac{d^dl}{(2\pi)^d} \, \frac{-1}{s - \varepsilon_{\bf l} - \varepsilon_{\bf P-l}}  .\label{eq:defL2}
\end{align}
The momentum integration diverges in $d=2,3$, which is canceled by the renormalization of the coupling constant $g$. For reference, we note the result for the $T$ matrix
\begin{align}
T_2^{-1}(s) &= \begin{cases}
\dfrac{m}{2 \hbar^2} \Bigl(- a + \sqrt{- \dfrac{\hbar^2}{m s}}\Bigr) & {\rm (1D)} \\[1.5ex]
\dfrac{m}{2\pi\hbar^2} \ln a \sqrt{- \dfrac{m}{\hbar^2} s} & {\rm (2D)} \\[1.5ex]
\dfrac{m}{4\pi\hbar^2} \Bigl(- a^{-1} + \sqrt{- \dfrac{m}{\hbar^2} s}\Bigr) & {\rm (3D)} ,
\end{cases} \label{eq:defT2}
\end{align}
where, by Galilean invariance, $T_2(s, {\bf P}) = T_2(s-\frac{1}{2}\varepsilon_{\bf P})$. Recall the structure of the imaginary part of the $T$ matrix, ${\rm Im} \, T_2(s+i0)$, which consists of a bound-state peak (where present) at the bound-state energy $E_b = - \frac{\hbar^2}{ma^2}$, and a continuum of scattering states for positive energies. It will be convenient at a later point to use dimensionless scattering matrices $\tilde{T}_2(s)$ defined via
\begin{align}
\tilde{T}_2(s) &= \frac{m \lambda_T^{2-d}}{\hbar^2 \Omega_d} T_2(s) , \label{eq:defT2tilde}
\end{align}
where $\Omega_d = 2 \pi^{d/2}/\Gamma(d/2)$ is the $d$-dimensional surface element, i.e., we separate a factor $\frac{2\hbar^2}{m\lambda_T}$, $\frac{2\pi\hbar^2}{m}$, and $\frac{4\pi\hbar^2\lambda_T}{m}$ in $d=1,2$, and $3$, respectively. 

For further reference, we note a useful representation of the contact to leading order ${\cal O}(z^2)$ in the fugacity:
\begin{align}
\lambda_T^4 C &= 4 \pi^2 z^2 \beta^2 \frac{2^{d/2}}{\lambda_T^d} \int_{-\infty}^\infty ds \, \frac{e^{-\beta s}}{\pi} \, {\rm Im} T_2(s+i0) . \label{eq:contact}
\end{align}
The integral can be evaluated in closed analytical form in 1D~\cite{hoffman15} and 3D. This expression will be used in Sec.~\ref{sec:III} when extracting the high-frequency tails of the viscosity spectral function, the magnitude of which is set by the contact.

Note that in this paper, we focus on the viscosity of Fermi gases, but all calculations apply with minor modifications to the Bose gas as well. The zero-range model for Bose gases replaces the Fermi by Bose fields in Eq.~\eqref{eq:zerorange} and conventionally sets $g \to g/4$. The scattering $T$ matrix in Eq.~\eqref{eq:defT} then has an additional symmetry factor of $1/2$ in front of the loop integral, and the overall Bose $T$ matrix is larger by a factor of $2$ compared to the Fermi case~\cite{barth15}. The inclusion of a three-body term in the zero-range model will not affect any results to leading order ${\cal O}(z^2)$ in the fugacity, which is fully determined by two-body correlations~\cite{barth15b}.

\subsection{Stress correlator}\label{sec:IIb}

\begin{figure}[t!]
\begin{center}
\subfigure[]
{
\raisebox{0.2cm}{\scalebox{0.35}{\includegraphics{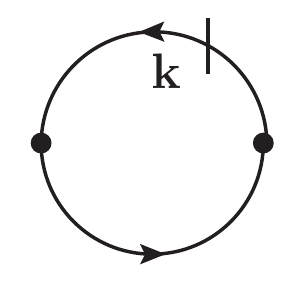}}}\quad
\raisebox{0.2cm}{\scalebox{0.35}{\includegraphics{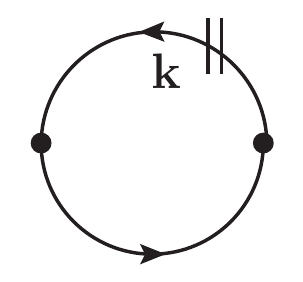}}}\quad
\raisebox{0.2cm}{\scalebox{0.35}{\includegraphics{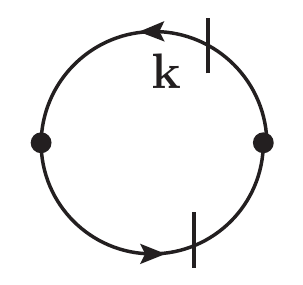}}}
}

\subfigure[]
{
\raisebox{0.2cm}{\scalebox{0.35}{\includegraphics{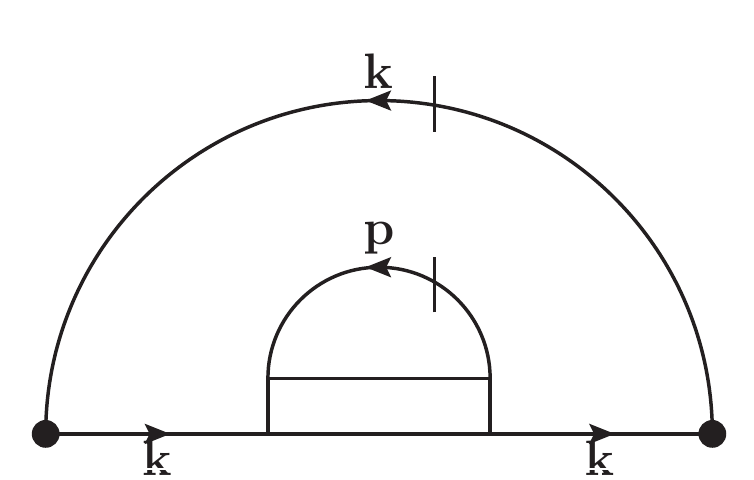}}}\quad
\scalebox{0.35}{\includegraphics{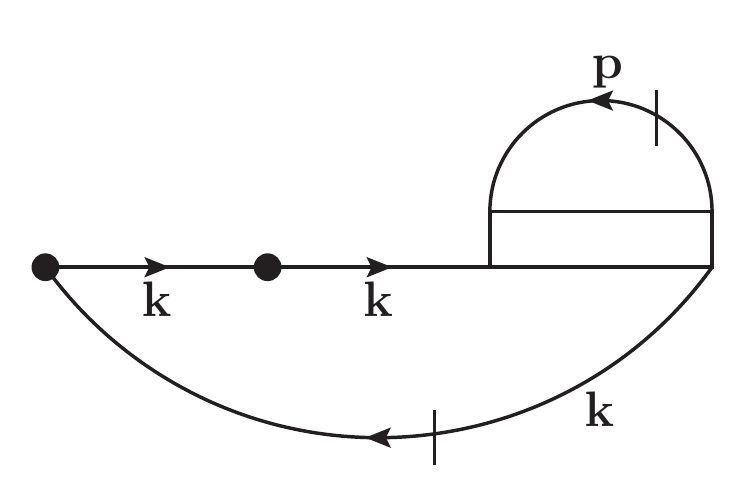}}
}
\subfigure[]
{
\raisebox{0.2cm}{\scalebox{0.35}{\includegraphics{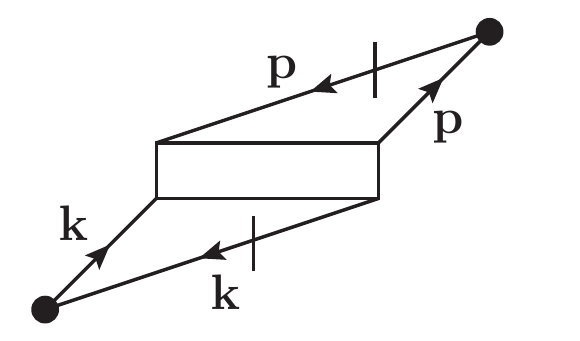}}}\quad
\scalebox{0.35}{\includegraphics{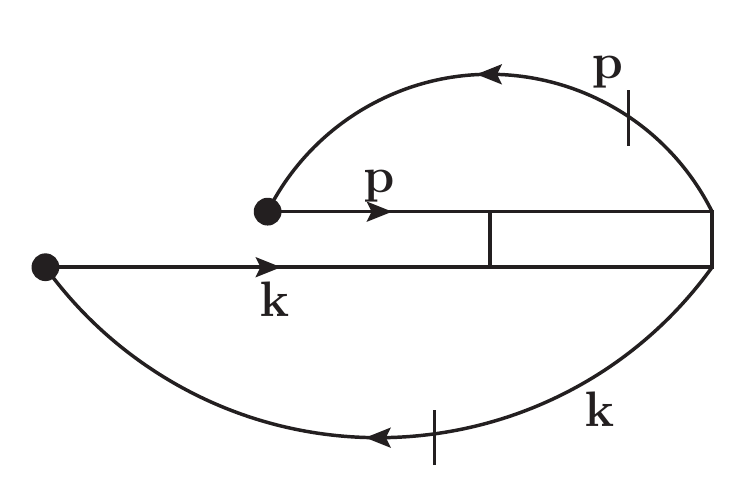}}
}
\subfigure[]
{
\scalebox{0.3}{\includegraphics{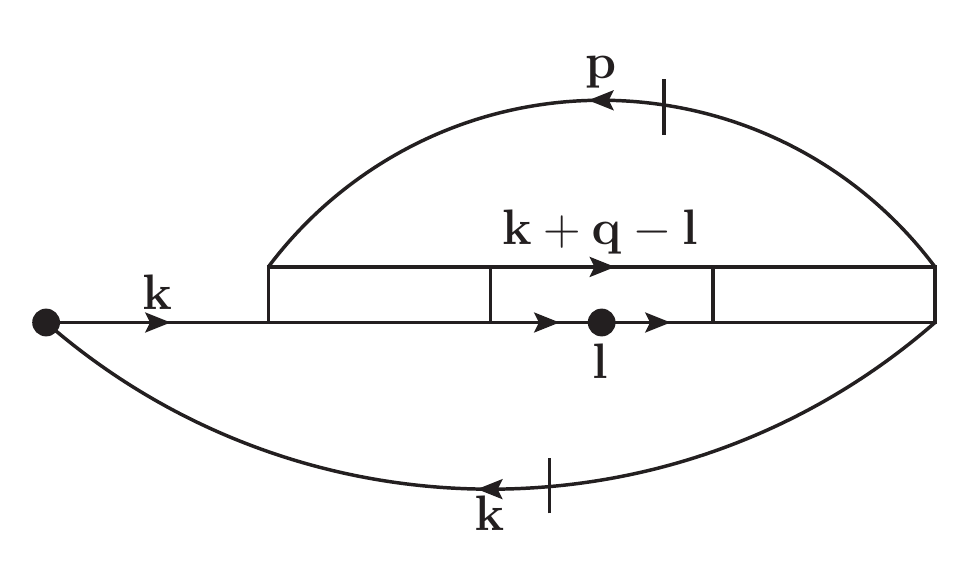}}\quad
\scalebox{0.3}{\includegraphics{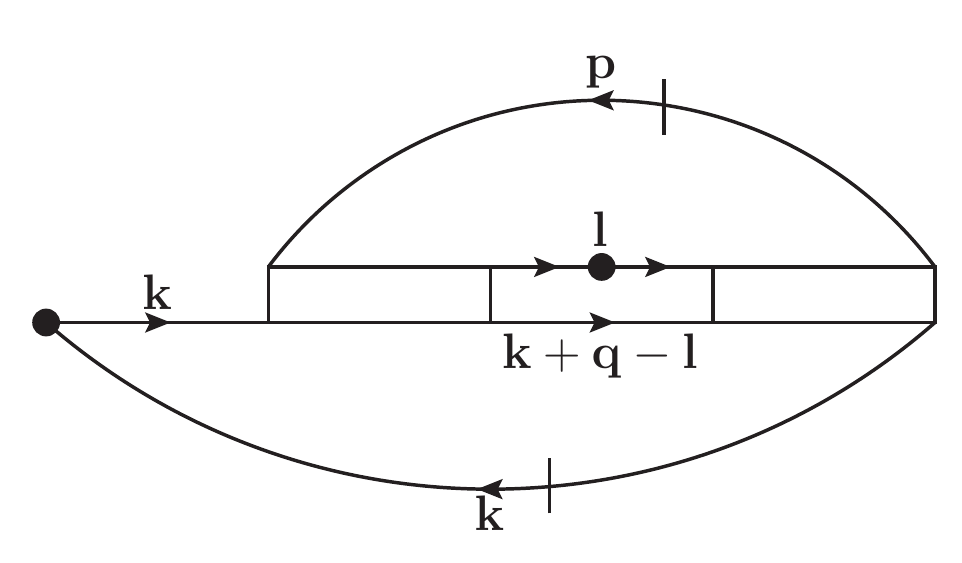}}
}
\caption{Feynman diagram representation of the shear viscosity stress correlator up to second order in the fugacity ${\cal O}(z^2)$. Slashed lines denote virial propagators, dots show the stress tensor insertions $T^0_{xy}(0)$ and $T^0_{xy}(\tau)$, and the box is the $T$ matrix. The diagrams are divided into (a) noninteracting, (b) self-energy, (c) Maki-Thompson, and (d) Aslamazov-Larkin diagrams.
 }
\label{fig:3}
\end{center}
\end{figure}

In this section we discuss the Feynman diagram expansion of the stress correlator $\chi_{xy,xy}^R(\omega)$ used in the computation of the shear viscosity~\eqref{eq:defshear} up to quadratic order in the fugacity ${\cal O}(z^2)$. A Feynman diagram representation of the imaginary-time correlator $\chi_{xy,xy}(\tau)$ is obtained by drawing all connected diagrams with a stress tensor insertion $\hat{\Pi}_{xy}$ at $\tau=0$ and at $0< \tau < \beta$. All such diagrams up to quadratic order are shown in Fig.~\ref{fig:3}. We then take the discrete Fourier transform to obtain the response as a function of Matsubara frequencies. The leading-order ${\cal O}(z)$ contribution [Fig.~\ref{fig:3}(a)] is the same as for noninteracting gas. Indeed, the noninteracting contribution can be summed to all orders in $z$. Using the Feynman rules introduced in the preceding section, we obtain the standard kinetic result
\begin{align}
&{\rm Im} \, \chi_{xy,xy}^{\rm R,ni}(\omega) = \delta(\beta \hbar \omega) (1 - e^{-\beta \hbar \omega}) \nonumber \\
&\qquad\times2 \int \frac{d^dk}{(2\pi)^d} \biggl(- \frac{\partial f(\varepsilon_{{\bf k}})}{\partial \varepsilon_{\bf k}}\biggr) T^0_{xy}({\bf k}) T^0_{xy}({\bf k}) , \label{eq:stressnonint}
\end{align}
where the prefactor $2$ accounts for the two spin species. This term will contribute a $\delta$ function to the shear viscosity spectral function $\eta(\omega)$.

Two-particle interactions enter at second order ${\cal O}(z^2)$. The Feynman diagrams that represent the interacting ${\cal O}(z^2)$ contribution to the shear viscosity spectral function~\eqref{eq:defshear} are shown in Figs.~\ref{fig:3}(b) and~\ref{fig:3}(c). These diagrams are commonly divided into three separate classes~\cite{larkin05}: (a) self-energy diagrams, which affect the spin-symmetric part of the response function and contain a self-energy insertion on one propagator; (b) Maki-Thompson diagrams, which contribute to the spin-antisymmetric response and contain a vertex correction; and (c) Aslamazov-Larkin diagrams. In the following, we compute each of these contributions separately, focusing on the main steps of the calculation. To illustrate the calculations in detail, Appendix~\ref{app:2} contains an in-depth discussion of the self-energy contribution.

\subsubsection{Self-energy}

There are two self-energy diagrams as shown in Fig.~\ref{fig:3}(b) Using the imaginary-time Feynman rules and applying the convolution theorem for the Laplace transform, the first one evaluates to
\begin{align}
&\chi_{xy,xy}^{\rm SE,a}(i \omega_n) = 2 z^2 \int \frac{d^dp}{(2\pi)^d} \int \frac{d^dk}{(2\pi)^d} \, T^0_{xy}({\bf k}) T^0_{xy}({\bf k}) \nonumber \\
&\times \int_{\rm Bw} \frac{ds}{2\pi i} \, e^{- \beta s} \, \frac{T_2(s + i\hbar\omega_n, {\bf k} + {\bf p})}{(s - \varepsilon_{{\bf k}} - \varepsilon_{{\bf p}}) (s + i\hbar\omega_n - \varepsilon_{{\bf k}} - \varepsilon_{{\bf p}})^2} . \label{eq:se_a}
\end{align}
The overall prefactor $2$ accounts for the spin summation. The Bromwich integral and the analytic continuation from Matsubara to real frequency $i\omega_n \to \omega + i0$ are performed as outlined in Appendix~\ref{app:1}. This gives the self-energy contribution to the retarded response:
\begin{align}
&\frac{{\rm Im} \, \chi_{xy,xy}^{\rm R,SE,a}(\omega)}{1-e^{-\beta \hbar \omega}} = - 2 z^2 \int \frac{d^dp}{(2\pi)^d} \int \frac{d^dk}{(2\pi)^d} \, e^{- \beta (\varepsilon_{\bf k} + \varepsilon_{\bf p})} \nonumber \\
&\qquad\times {\rm Im} \biggl[\frac{T_2(\hbar \omega + i0 + \frac{1}{2} \varepsilon_{\bf k-p} 
)}{(\hbar \omega + i0)^2}\biggr] T^0_{xy}({\bf k}) T^0_{xy}({\bf k}) . \label{eq:res_chiSEa}
\end{align}
The second self-energy diagram in Fig.~\ref{fig:3}(b) contributes an identical term up to the sign with $\omega \to -\omega$:
\begin{align}
{\rm Im} \, \chi_{xy,xy}^{\rm R,SE,b}(\omega) &= - {\rm Im} \, \chi_{xy,xy}^{\rm R,SE,a}(-\omega) . \label{eq:res_chiSEb}
\end{align}
The imaginary part of the self-energy contribution is antisymmetric in the frequency, as expected for a retarded response function. Appendix~\ref{app:2} contains a detailed discussion of the above derivation.

\subsubsection{Maki-Thompson}

The Maki-Thompson diagrams are shown in Fig.~\ref{fig:3}(c). The evaluation proceeds in a similar way as for the self-energy part. For the first diagram, we obtain
\begin{align}
&\frac{{\rm Im} \, \chi_{xy,xy}^{\rm R,MT,a}(\omega)}{1-e^{-\beta \hbar \omega}} = - 2 z^2 \int \frac{d^dp}{(2\pi)^d} \int \frac{d^dk}{(2\pi)^d} \, e^{-\beta (\varepsilon_{\bf k} + \varepsilon_{\bf p})} \nonumber \\
&\qquad\times {\rm Im} \biggl[\frac{T_2(\hbar \omega + i0 + \frac{1}{2} \varepsilon_{\bf k-p}
)}{(\hbar \omega + i 0)^2}\biggr] T^0_{ij}({\bf k}) T^0_{kl}({\bf p}) .
\end{align}
Again, the second diagram in Fig.~\ref{fig:3}(c) contributes a term of equal magnitude with the replacement $\omega \to -\omega$:
\begin{align}
{\rm Im} \, \chi_{xy,xy}^{\rm R,MT,b}(\omega) &= - {\rm Im} \, \chi_{xy,xy}^{\rm R,MT,a}(-\omega) .
\end{align}

\subsubsection{Aslamazov-Larkin}

The computation of the Aslamazov-Larkin diagram is more involved compared to self-energy and Maki-Thompson contributions. Corresponding Feynman diagrams are shown in Fig.~\ref{fig:3}(d). They evaluate to
\begin{align}
&\chi_{xy,xy}^{\rm AL}(i \omega_n) = 4 z^2 \int \frac{d^dp}{(2\pi)^d} \int \frac{d^dk}{(2\pi)^d}  \, T_{xy}^0({\bf k}) \nonumber \\
&\times\int_{\rm Bw} \frac{ds}{2\pi i} \, e^{- \beta s} \, \frac{T_2(s, {\bf k} + {\bf p}) T_2(s + i\hbar\omega_n, {\bf k} + {\bf p})}{(s - \varepsilon_{\bf k} - \varepsilon_{\bf p}) (s + i\hbar\omega_n - \varepsilon_{{\bf k}} - \varepsilon_{\bf p})} \nonumber \\
&\times \int \frac{d^dl}{(2\pi)^d} \, \frac{T_{xy}^0({\bf l})}{(s - \varepsilon_{\bf l} - \varepsilon_{\bf k + p - l}) (s + i\hbar\omega_n - \varepsilon_{\bf l} - \varepsilon_{\bf k + p - l})} . \label{eq:AL}
\end{align}
There are two distinct diagrams in Fig.~\ref{fig:3}(d) and two spin species, hence the overall prefactor $4$. The integral in the last line of Eq.~\eqref{eq:AL} corresponds to the internal three-propagator loop at the center of the diagram (sometimes called a triangle diagram). We evaluate the integral using Eq.~\eqref{eq:defT}:
\begin{align}
&\int \frac{d^dl}{(2\pi)^d} \, \frac{T_{xy}^0({\bf l})}{(s - \varepsilon_{\bf l} - \varepsilon_{\bf k + p - l}) (s + i\hbar\omega_n - \varepsilon_{\bf l} - \varepsilon_{\bf k + p - l})} \nonumber \\
&= \frac{T_{xy}^0({\bf k} + {\bf p})}{4 (i\hbar\omega_n)} \Bigl[T_2^{-1}(s, {\bf k + p}) - T_2^{-1}(s + i\hbar\omega_n, {\bf k + p})\Bigr]  . \label{eq:ward}
\end{align}
Note that Eq.~\eqref{eq:ward} is a Ward identity which holds for the full $T$ matrix.  Furthermore, we simplify the product of the two $T$-matrices using
\begin{align}
&T_2(s, {\bf k+p}) T_2(s + i\hbar\omega_n, {\bf k+p}) \nonumber \\
&\quad = \frac{T_2(s + i\hbar\omega_n, {\bf k+p}) - T_2(s, {\bf k+p})}{T_2^{-1}(s, {\bf k+p}) - T_2^{-1}(s + i\hbar\omega_n, {\bf k+p})} .
\end{align}
The diagram now takes a simpler form
\begin{align}
&\frac{{\rm Im} \, \chi_{xy,xy}^{\rm R,AL,a}(\omega)}{1 - e^{-\beta \hbar \omega}} = - z^2 \int \frac{d^dp}{(2\pi)^d} \int \frac{d^dk}{(2\pi)^d} \, e^{- \beta (\varepsilon_{\bf k} + \varepsilon_{\bf p})} \nonumber \\
&\qquad\times {\rm Im} \Bigl[\frac{T_2(\hbar \omega + i 0 + \frac{1}{2} \varepsilon_{\bf k-p}
)}{(\hbar \omega + i0)^2}\Bigr] T_{xy}^0({\bf k}) T_{xy}^0({\bf k} + {\bf p}) ,
\end{align}
and we split off a second term
\begin{align}
{\rm Im} \, \chi_{xy,xy}^{\rm R,AL,b}(\omega) = - {\rm Im} \, \chi_{xy,xy}^{\rm R,AL,a}(- \omega) .
\end{align}

\subsection{Contact correlator}\label{sec:IIc}

\begin{figure}[t!]
\begin{center}
\raisebox{0.2cm}{\scalebox{0.35}{\includegraphics{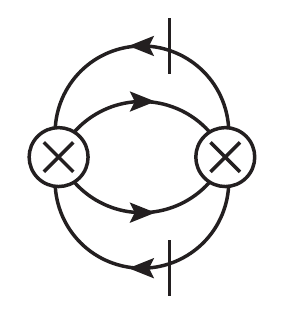}}}
\scalebox{0.32}{\includegraphics{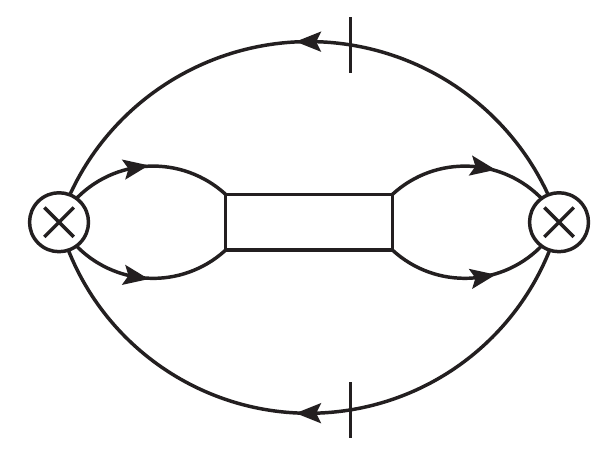}}\quad
\scalebox{0.32}{\includegraphics{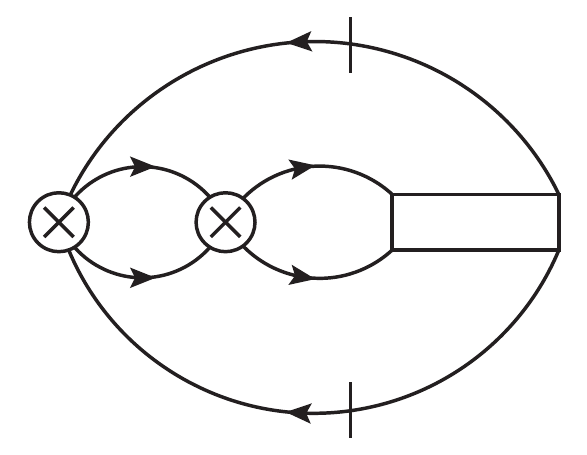}}\quad
\scalebox{0.3}{\includegraphics{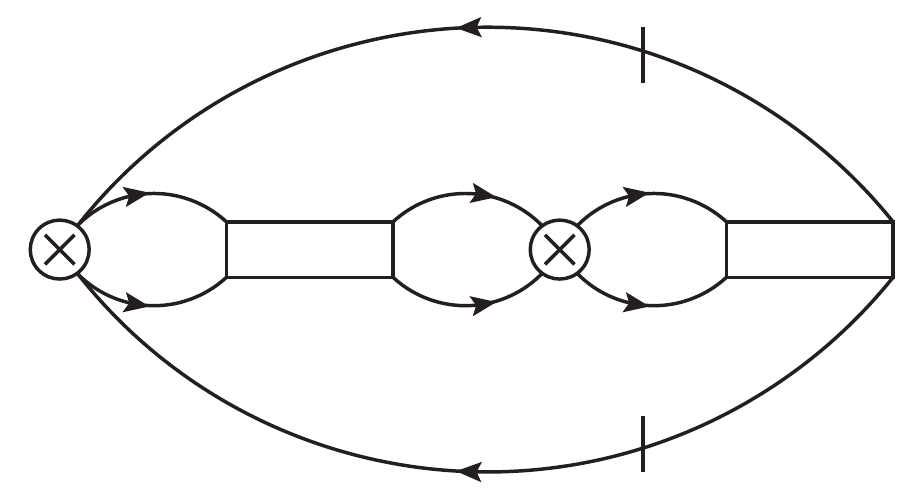}}
\caption{Feynman diagram representation of the bulk viscosity contact correlator up to second order in the fugacity ${\cal O}(z^2)$. Slashed lines denote virial propagators, dots show the contact operator insertions $\hat{C}(0)$ and $\hat{C}(\tau)$, and the box is the $T$ matrix.}
\label{fig:4}
\end{center}
\end{figure}

We now turn to the contact correlator $\chi_{C,C}(\omega)$ used in the computation of the bulk viscosity spectral function. The leading-order contribution to the correlation function is of second order in the fugacity ${\cal O}(z^2)$. Corresponding diagrams are shown in Fig.~\ref{fig:4}. They involve the $T$ matrix coefficients and the bare two-particle propagator $L_{\bf P}(\tau)$ [Eq.~\eqref{eq:defL1}] as subblocks. The calculation proceeds in a similar way as in the preceding section. After analytic continuation, the sum of all four diagrams is
\begin{align}
&\chi_{C,C}(\omega) = \frac{m^4 z^2 g^4}{\hbar^8} \int \frac{d^dP}{(2\pi)^d} \int_{-\infty}^\infty \frac{ds}{\pi } \, e^{- \beta s} \, \nonumber \\
&\times{\rm Im} \Bigl[L_{\bf P}(s+\omega+i0) \nonumber \\
& \times \Bigl(1+ T_2(s+\omega+i0, {\bf P})L_{\bf P}(s+\omega+i0)\Bigr)\Bigr] \nonumber \\ 
& \times {\rm Im} \Bigl[L_{\bf P}(s+i0) \Bigl(1+ T_2(s+i0, {\bf P})L_{\bf P}(s+i0)\Bigr)\Bigr] , \label{eq:concor}
\end{align}
where $L_{\bf P}(s)$ is the Laplace transform of the bare two-particle propagator defined in Eq.~\eqref{eq:defL2}. Equation~\eqref{eq:concor} is simplified using the definition of the $T$ matrix~\eqref{eq:defT}, which gives
\begin{align}
&\frac{\chi_{C,C}(\omega)}{1 - e^{-\beta \hbar \omega}} = \frac{m^4 z^2}{\hbar^8} \frac{2^{d/2}}{\lambda_T^d} \nonumber \\
& \times \int_{-\infty}^\infty \frac{ds}{\pi } \, e^{- \beta s} \, {\rm Im} \bigl[T_2(s+i0)\bigr] {\rm Im} \bigl[T_2(s+\omega+i0)\bigr] . \label{eq:contactcorrelator}
\end{align}
This expression is antisymmetric in $\omega$ as can be seen by shifting the integration variable $s \to s - \omega$. The integral~\eqref{eq:contactcorrelator} is convenient for a direct numerical evaluation.

\section{Results}\label{sec:III}

In this section we combine the results of the preceding section to compute in turn the shear viscosity and bulk viscosity spectral functions in Secs.~\ref{sec:IIIa} and~\ref{sec:IIIb}, respectively. We present integral representations of the spectra, from which the general three-part structure of the spectral function as shown in Fig.~\ref{fig:1}---a Drude or dimer zero-frequency peak, a broad free-particle continuum, and a bound-state contribution above threshold---becomes apparent. Numerical results for the spectral functions are presented for various dimensions and scattering lengths. Universal high-frequency tails and sum rules are obeyed in our calculations, which demonstrates the internal consistency of the virial expansion up to this order. We discuss the static limit in some detail and make connections with kinetic theory.

\subsection{Shear viscosity}\label{sec:IIIa}

\begin{figure}[t]
\scalebox{0.9}{\includegraphics{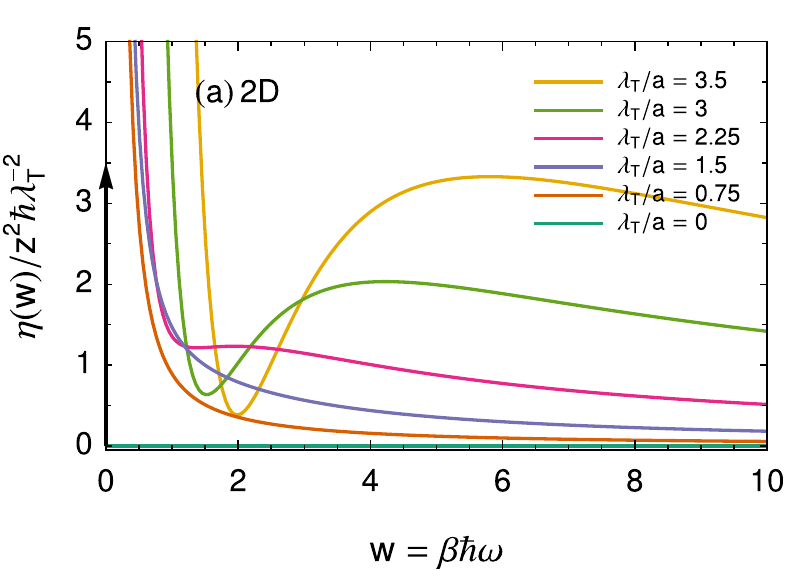}\qquad}
\scalebox{0.9}{\includegraphics{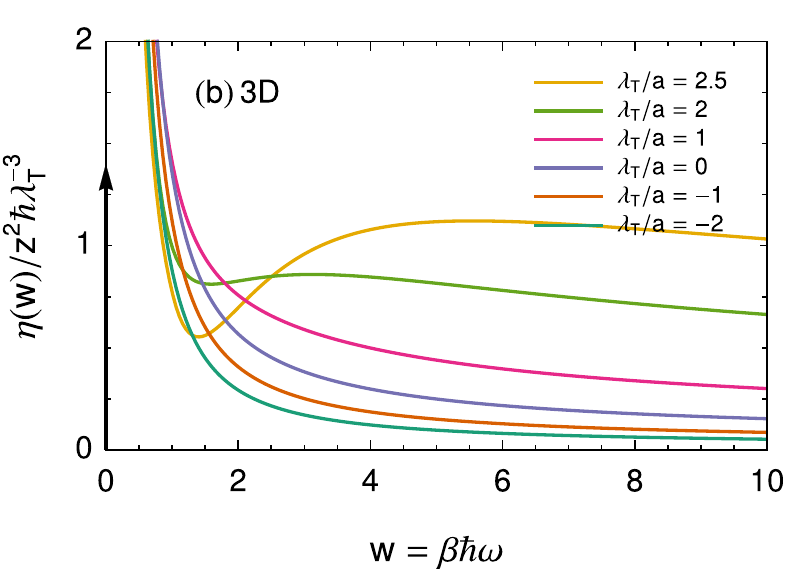}\qquad}
\caption{Shear viscosity spectral function in (a) 2D and (b) 3D for different values of the inverse scattering length $\lambda_T/a$.  The spectral function has a universal power-law high-frequency tail [Eq.~\eqref{eq:etahigh}] and diverges as a power law at low frequencies. Above a threshold frequency, there is a bound-state contribution which dominates the spectral function in the dimer limit. The black arrow indicates the $\delta$-function Drude peak.}
\label{fig:5}
\end{figure}

We begin by computing the shear viscosity spectral function. The zero-frequency Drude peak has a leading noninteracting ${\cal O}(z)$ contribution [Eq.~\eqref{eq:stressnonint}]. At finite frequency, the spectral function is ${\cal O}(z^2)$ and given by the sum of self-energy, Maki-Thompson, and Aslamazov-Larkin contribution
\begin{align}
&\eta (\omega) = \frac{{\rm Im} \, [\chi_{xy,xy}^{\rm R,SE}(\omega)+\chi_{xy,xy}^{\rm R,MT}(\omega)+\chi_{xy,xy}^{\rm R,AL}(\omega)]}{\omega} \nonumber \\
&= \frac{z^2}{4} \frac{(1-e^{-\beta \hbar \omega})}{\hbar^2 \omega^3} \int \frac{d^dp}{(2\pi)^d} \int \frac{d^dk}{(2\pi)^d} \, e^{- \beta (\varepsilon_{\bf k} + \varepsilon_{\bf p})} \,   \nonumber \\
&\quad\times{\rm Im} \Bigl[T_2(\hbar \omega + i0 + \frac{1}{2} \varepsilon_{\bf k-p})\Bigr] T^0_{xy}({\bf k-p}) T^0_{xy}({\bf k} - {\bf p}) \nonumber \\
&\quad+ (\omega \to - \omega) , \label{eq:eta}
\end{align}
where we have symmetrized the stress tensor matrix element $T_{xy}^0$. For a numerical evaluation, it is convenient to transform to relative and center-of-mass momenta ${\bf P} = ({\bf k} + {\bf p})/2$ and ${\bf q} = {\bf k} - {\bf p}$, which restricts the angular integration to the matrix elements $T^0_{xy}({\bf q}) T^0_{xy}({\bf q})$. The center-of-mass momentum integration is then performed directly. 

The imaginary part of the scattering $T$ matrix $T_2(s+i0)$ [Eq.~\eqref{eq:defT2}] has a continuous spectrum at positive energies and (if present) a state pole at the bound-state energy $E_b = - \frac{\hbar^2}{ma^2}$. From the structure of the integrand in Eq.~\eqref{eq:eta}, we deduce the general form of the shear viscosity spectral function shown in Fig.~\ref{fig:1}(a). First, the free-particle part of the $T$ matrix contributes at all frequencies and forms the continuum indicated in the figure. Second, if the frequency exceeds the dimer binding energy, the second term in~\eqref{eq:eta} gives rise to a bound-state contribution with a threshold behavior $\sim (\hbar \omega + E_b)^{(d+2)/2}$, which is nonanalytic in 3D. At small frequencies, only the continuum part contributes to the spectrum. As is apparent from the integral~\eqref{eq:eta}, the spectral function diverges as $1/\omega^2$ in this limit. This divergence will be discussed in more detail at the end of the section. Figure~\ref{fig:5} shows numerical results for the shear viscosity spectral function in two and three dimensions for various values of the dimensionless inverse scattering length $\lambda_T/a$. The general structure of Fig.~\ref{fig:1}(a) is clearly reflected in these results.

\subsubsection{High-frequency tails}

The shear viscosity spectral function has a power-law high-frequency tail with a magnitude set by the contact~\cite{taylor10, hofmann11,goldberger12}. The leading-order term at large frequencies stems from the second term in Eq.~\eqref{eq:eta} with $\omega \to -\omega$. Transforming to a relative-energy integration variable $s=2 \varepsilon_{\bf q}$, we shift $s \to s + \omega$ and extend the $s$ integration to run from $-\infty$ to $+\infty$. Separating the leading power in $\hbar \omega$ from the integrand, we recognize the remaining integral as the leading term in the virial expansion of the contact [Eq.~\eqref{eq:contact}]. This gives
\begin{align}
\lim_{\omega \to \infty} \eta(\omega) &= 
\begin{cases}
\dfrac{\hbar^2 C}{8 m\omega} & {\rm (2D)} \\[2ex]
\dfrac{\hbar^{3/2} C}{15 \pi \sqrt{m\omega}} & {\rm (3D)} .
\end{cases} \label{eq:etahigh}
\end{align}
These high-frequency tails agree with general results obtained using the operator product expansion (OPE) in 3D~\cite{hofmann11,goldberger12} and 2D~\cite{hofmann11}.

In the three-dimensional unitary limit, we can compare with a Keldysh calculation~\cite{enss11}, which was performed using a finite lifetime $\gamma/2$ in the single-particle self-energy. Here, $C = 16 \pi z^2/\lambda_T^4 = 4 m^2 z^2/\pi \beta^2 \hbar^4$ and we state the separate contributions of the self-energy, Maki-Thompson, and Aslamazov-Larkin terms to the high-frequency tail
\begin{align}
\lim_{\omega \to \infty} \eta_{\rm unitary}(\omega) &= \dfrac{4 m^2 z^2}{\pi \beta^2 \hbar^4} \frac{1+1+0}{30} \dfrac{\hbar^{3/2}}{\pi \sqrt{m \omega}} ,
\end{align}
which agrees with Ref.~\cite{enss11}. This indicates that the set of diagrams considered in Ref.~\cite{enss11} contains all the diagrams in Fig.~\ref{fig:3} that yield the correct high-temperature limit.

\subsubsection{Static shear viscosity and memory function approach}

Our results for the shear viscosity spectral function diverge at small frequencies such that a direct extrapolation to the static limit is not possible. In the following we discuss how this divergence is resummed using the memory function formalism~\cite{goetze72,forster90,enss11}, which yields a finite static shear viscosity that reproduces the kinetic theory result.

Hydrodynamics dictates a low-frequency form of the retarded response~\cite{forster90}
\begin{align}
\frac{\chi(\omega)}{\omega} &= \frac{W}{\omega - i/\tau} = W \frac{\omega \tau^2}{1+(\omega\tau)^2} +  W \frac{i \tau}{1+(\omega\tau)^2} , \label{eq:chihydro}
\end{align}
which has a finite limit for $\omega \to 0$. The static shear viscosity $\eta = \lim_{\omega \to 0} \frac{{\rm Im} \, \chi(\omega)}{\omega} = W \tau$ is then expressed as the ratio of a Drude weight $W$ and a viscous scattering rate $\tau^{-1}$. The scattering rate $\tau^{-1}$, which vanishes for noninteracting systems, admits a separate fugacity expansion. A direct virial expansion of the correlation function thus applies in the limit $\omega \tau \gg 1$ and does not connect with the hydrodynamic expression~\eqref{eq:chihydro}. This is the origin of the $1/\omega^2$ divergence of the shear viscosity spectral function~\eqref{eq:eta} for $\beta \hbar \omega \ll 1$.

Progress can be made using the memory function formalism~\cite{goetze72,forster90,enss11}, which instead of the full correlation function proposes to expand the memory function $M(\omega)$ defined through
\begin{align}
\frac{\chi(\omega)}{\omega} &= \frac{\eta_0}{\omega - M(\omega)} , \label{eq:memory}
\end{align}
with $\eta_0$ a real constant. Note that the relation between the correlation function and the memory function is similar to the relation between the single-particle Green's function and self-energy. Since $\chi(-\omega) = \chi^*(\omega)$, we have $M(-\omega) = - M^*(\omega)$. Using the low-frequency expansion $M(\omega) = i M_0 + \omega M_1 + {\cal O}(\omega^2)$ with $M_0$ and $M_1$ real, we find
\begin{align}
W &= \frac{\eta_0}{1-M_1},  \quad
\frac{1}{\tau} = \frac{M_0}{1-M_1} ,
\end{align}
which implies, for the static limit of the viscosity spectral function,
\begin{align}
\eta = W \tau = \frac{\eta_0}{M_0} . \label{eq:memory2}
\end{align}
We can extract the coefficients $\eta_0$ and $M_0$ from the virial expansion results by matching the low-frequency terms of the virial expansion with the expansion of Eq.~\eqref{eq:memory},
\begin{align}
\frac{\chi(\omega)}{\omega}  &= \frac{\eta_0}{\omega} + \frac{\eta_0}{\omega^2} M(\omega) + \cdots , \label{eq:memory3}
\end{align}
In this way, we obtain a static limit of the shear viscosity using~\eqref{eq:memory2}. Intuitively, the memory kernel result for the static viscosity~\eqref{eq:memory2} is obtained from a geometric resummation of the leading two low-frequency terms in Eq.~\eqref{eq:memory3}. Note that a similar approach has been used to determine the line shift in the dynamic structure factor of interacting quantum gases~\cite{hofmann17}. 

To be definite, we consider the three-dimensional case in the following. First from Eq.~\eqref{eq:stressnonint} we obtain $\eta_0$,
\begin{align}
\eta_0 &= 2 \int \frac{d^3k}{(2\pi)^3} \biggl(- \frac{\partial f(\varepsilon_{{\bf k}})}{\partial \varepsilon_{\bf k}}\biggr) [T^0_{xy}({\bf k})]^2 \nonumber \\
&= \frac{(2m)^{3/2} z}{2 \pi^{3/2} \hbar^3 \beta^{5/2}} + {\cal O}(z^2) , \label{eq:eta0}
\end{align}
valid to leading order in the fugacity. Second, the coefficient $\eta_0 M_0$ follows from the low-frequency divergence of Eq.~\eqref{eq:eta}, which is obtained by expanding the prefactor and setting $\omega=0$ in the integrand. While at first sight this expression might seem unwieldy, we will show that it takes the form of a (linearized) collision integral such that $M_0$ can be identified with a viscous scattering time $\tau_\eta^{-1}$. In this way, using Eq.~\eqref{eq:memory2}, the virial expansion combined with the memory kernel approach reproduces the calculation of the viscosity within kinetic theory.

Within kinetic theory, the static shear viscosity takes the form~\cite{massignan05,bruun05,bruun07}
\begin{align}
\eta &= 2 \tau_{\eta} \int \frac{d^3p}{(2\pi)^3} [T_{xy}^0({\bf k})]^2 \biggl(-\frac{\partial f_\sigma}{\partial \varepsilon_{\bf k}}\biggr) = \eta_0 \tau_\eta ,
\end{align}
where we use the definition ion of $\eta_0$ in Eq.~\eqref{eq:eta0} and following the literature we introduce a viscous relaxation time $\tau_{\eta}$~\cite{massignan05,bruun05,bruun07},
\begin{align}
\frac{1}{\tau_{\eta}} &= \frac{1}{{\cal N}} \int \frac{d^3p}{(2\pi)^3} \, \int \frac{d^3k}{(2\pi)^3} \, \int d\Omega \, \frac{d\sigma}{d\Omega} \, |{\bf v}_{\rm rel}| \frac{\beta^2 \hbar^2}{m^2} q_x^2 q_y^2 \nonumber \\
&\times f(\varepsilon_{\bf p}) \, f(\varepsilon_{{\bf k}}) \, [1 - f(\varepsilon_{{\bf p}'})] \, [1 - f(\varepsilon_{{\bf k}'})] , \label{eq:kinetic}
\end{align}
with a normalization factor
\begin{align}
{\cal N} &= 2 \beta \int \frac{d^3k}{(2\pi)^3} \biggl(- \frac{\partial f(\varepsilon_{{\bf k}})}{\partial \varepsilon_{\bf k}}\biggr) [T_{xy}^0({\bf k})]^2  = \beta \eta_0 ,
\end{align}
which has dimension of inverse volume. In Eq.~\eqref{eq:kinetic}, $\frac{d\sigma}{d\Omega}$ is the differential scattering cross section for the scattering of two particles with opposite spin and wave vectors ${\bf k}$ and ${\bf p}$ to states ${\bf k}'$ and ${\bf p}'$, respectively, and ${\bf v}_{\rm rel} = \frac{2\hbar{\bf q}}{m}$ is the relative velocity. 

We now show that $\tau_\eta^{-1}$, given in Eq.~\eqref{eq:kinetic}, and $M_0$, given by Eq.~\eqref{eq:eta}, are identical at high temperatures, i.e., $\lim_{T\to \infty} \tau_\eta^{-1} = M_0$. At high temperatures, Pauli blocking is not effective and the Fermi factors in Eq.~\eqref{eq:kinetic} reduce to a Maxwell-Boltzmann distribution
\begin{align}
f(\varepsilon_{\bf p}) \, f(\varepsilon_{{\bf k}}) \, [1 - f(\varepsilon_{{\bf p}'})] \, [1 - f(\varepsilon_{{\bf k}'})] &= z^2 e^{-\beta (\varepsilon_{\bf p} + \varepsilon_{{\bf k}})} .
\end{align}
Furthermore, the optical theorem relates the $s$-wave scattering cross section to the imaginary part of the $T$ matrix
\begin{align}
\frac{d\sigma}{d\Omega} \, |{\bf v}_{\rm rel}| &= \frac{1}{2\pi\hbar} {\rm Im} \, T_2(2 \varepsilon_{\bf q}) . 
\end{align}
The equality between $\tau_\eta^{-1}$ and $M_0$ then follows directly by comparing with Eq.~\eqref{eq:eta}. From Eq.~\eqref{eq:memory2} we then obtain the static shear viscosity. Our calculations provide a rigorous correspondence between the high-temperature limit and kinetic theory for all scattering lengths.

A similar result was obtained for the unitary Fermi gas~\cite{enss11}. We note an additional check of our results for the shear viscosity spectral function in the unitarity limit: Writing the divergent contribution of the self-energy, Maki-Thompson, and Aslamazov-Larkin terms separately, we find
\begin{align}
\lim_{\omega\to0} \eta_{\rm unitary}(\omega) &= \dfrac{z^2 (2m)^{3/2}}{15 \pi^{5/2} \beta^{7/2} \hbar^4 \omega^2} \dfrac{43+3-30}{2 \sqrt{2}} .
\end{align}
Our result agrees with Ref.~\cite{enss11} if we identify $\hbar \omega \to 2 i \gamma$. 

\subsection{Bulk viscosity}\label{sec:IIIb}

\begin{figure}[t]
\scalebox{0.9}{\includegraphics{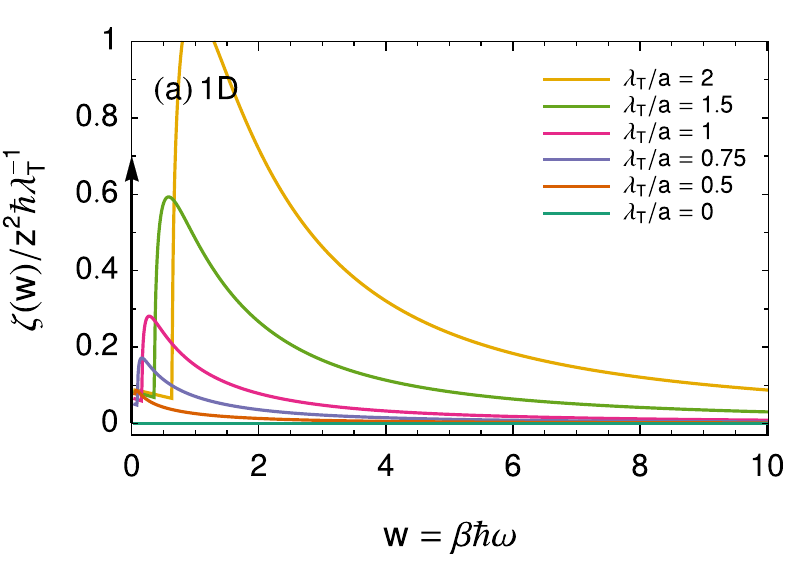}\qquad}
\scalebox{0.9}{\includegraphics{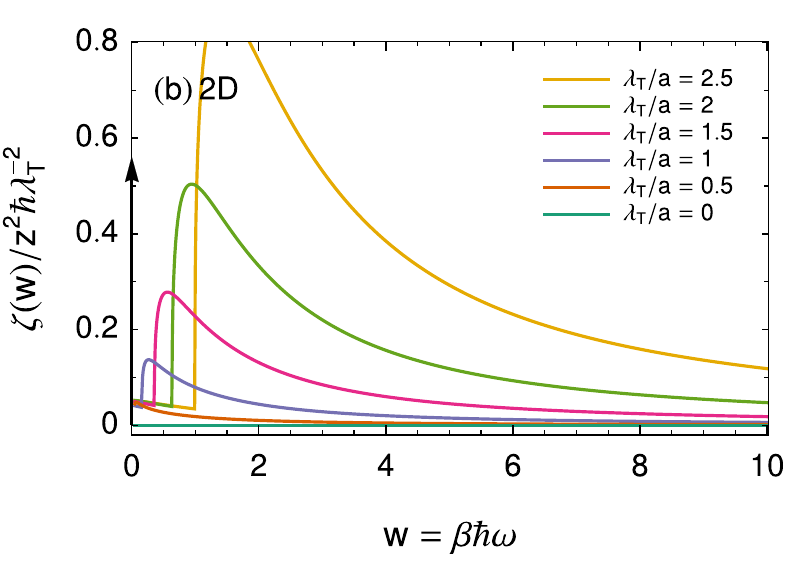}\qquad}
\scalebox{0.88}{\includegraphics{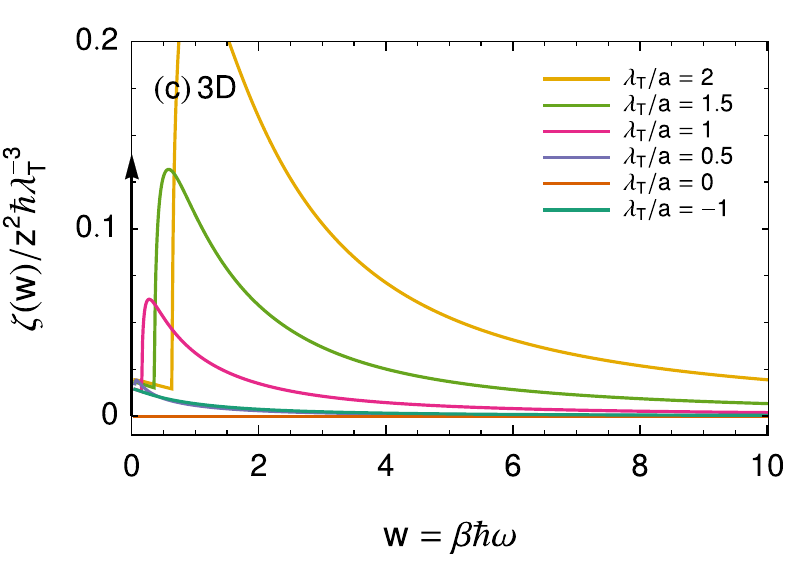}\qquad}
\caption{Bulk viscosity spectral function in (a) 1D, (b) 2D, and (c) 3D for different values of the dimensionless inverse scattering length $\lambda_T/a$.  The spectral function has a universal power-law high-frequency tail, Eq.~\eqref{eq:bulkhigh}. As for the shear viscosity, there is a bound state contribution which dominates the spectral function in the dimer limit above a threshold frequency.}
\label{fig:6}
\end{figure}

This section presents results for the bulk viscosity spectral function $\zeta(\omega)$. From the definitions~\eqref{eq:defbulkbetter} and~\eqref{eq:defO} and our result for the contact correlation function~\eqref{eq:contactcorrelator}, we obtain the integral representation of the spectral function:
\begin{align}
&\frac{\lambda_T^d}{\hbar z^2} \zeta(\omega) = \frac{1 - e^{-\beta \hbar \omega}}{d^2 \hbar \omega} \biggl(\frac{\lambda_T^2}{a^2}\biggr)^{d-2} \nonumber \\
& \times 2^{d/2} \int_{-\infty}^\infty \frac{ds}{\pi } \, e^{- \beta s} \, {\rm Im} \bigl[\tilde{T}_2(s+i0)\bigr] {\rm Im} \bigl[\tilde{T}_2(s+\hbar\omega+i0)\bigr] , \label{eq:zeta}
\end{align}
where the dimensionless scattering matrix $\tilde{T}_2$ is defined in Eq.~\eqref{eq:defT2tilde}. This expression is a symmetric function of the frequency $\omega$. In the following we discuss the structure of the spectral function and provide numerical results. We also discuss the high-frequency tail and sum rules and obtain analytical results for the static limit.

From the structure of the integrand in Eq.~\eqref{eq:zeta}, we directly infer the general form of the spectrum shown in Fig.~\ref{fig:1}(b). The first contribution arises from the integration over the free spectrum at positive energies. This part contributes at all frequencies and forms a continuous band. The integral in Eq.~\eqref{eq:zeta} is finite at $\omega=0$; hence the spectral function approaches a constant value for small frequencies. The second contribution appears for frequencies larger than the bound-state energy, for which we pick up the pole of the first scattering matrix. Notably, it has a nonanalytic frequency dependence right above the threshold energy which follows from the nonanalytic form of the $T$ matrix~\eqref{eq:defT2}. Finally, a third contribution at $\omega = 0$ arises from the pinching of the bound-state poles in the integrand~\eqref{eq:zeta}. This part gives rise to a $\delta$-function contribution $\sim \delta(\omega)$. The bound-state and $\delta$ contributions can be evaluated in closed analytical form.
		
Numerical results for the bulk viscosity spectral function as a function of the dimensionless inverse scattering length $\lambda_T/a$  are shown in Fig.~\ref{fig:6} for all dimensions and various values of the scattering length. The entire spectral function vanishes identically for $a\to\infty$. This point marks the unitary or noninteracting scale-invariant limit, where the bulk viscosity is expected to vanish on general grounds~\cite{son07}. Away from this limit, we clearly distinguish the continuum and the bound-state part of the spectral function. As is apparent from the figure, the bound-state contribution, when present, vastly dominates the continuum part. This observation is consistent with results for the zero-temperature spectral function~\cite{taylor12}.

\subsubsection{High-frequency tails and sum rules}

The high-frequency tail of the bulk viscosity spectral function is determined by the nonanalytic short-time behavior of the quantum gas and is expected to decay as a power law~\cite{taylor10, hofmann11,goldberger12}. Imposing a simultaneous scaling form, where the scattering length is assumed to scale with frequency, i.e., the dimensionless scaling variable $a \sqrt{\frac{m \omega}{\hbar}}$ is kept fixed, the dependence on the integration variable of the second term in the integrand of Eq.~\eqref{eq:zeta} can be neglected. The remaining integral is  identified with the contact expectation value at this order in the virial expansion [Eq.~\eqref{eq:contact}]. We obtain the high-frequency tails
\begin{align}
\lim_{\omega \to \infty} \zeta(\omega) &= 
\begin{cases}
\dfrac{\hbar^{5/2} C}{2 (m\omega)^{3/2}} \dfrac{1}{(a^{-1}\sqrt{\hbar/m\omega})^2 + 1} & {\rm (1D)} \\[2.5ex]
\dfrac{\hbar^{2} C}{4 m\omega} \dfrac{1}{\pi^2+ \ln^2 (a^{-1}\sqrt{\hbar/m\omega})^2} & {\rm (2D)} \\[2.5ex]
\dfrac{\hbar^{3/2} C}{36 \pi \sqrt{m\omega}} \dfrac{(a^{-1}\sqrt{\hbar/m\omega})^2}{(a^{-1}\sqrt{\hbar/m\omega})^2 + 1} & {\rm (3D)} .
\end{cases} \label{eq:bulkhigh}
\end{align}
These results agree with the OPE result in 3D~\cite{hofmann11,goldberger12} and the calculations in 2D~\cite{hofmann11}. The analytic high-frequency tails match our numerical calculations of the full bulk viscosity spectral function, and provide a check of our results. 

Unlike the shear viscosity spectral function, the bulk viscosity is finite for all nonzero frequencies, and hence so is the total spectral weight. An additional strong check of our result is provided by computing sum rules for the total spectral weight, which are related to the derivative of the contact~\cite{taylor10,hofmann11,goldberger12,taylor12}
\begin{align}
\frac{1}{\pi} \int_0^\infty d\omega \, \zeta(\omega) &=
\begin{cases}
 \dfrac{\hbar^2}{16 \pi m a} \dfrac{\partial C}{\partial a^{-1}} & {\rm (2D)} \\[2.5ex]
 \dfrac{\hbar^2}{72 \pi m a^2} \dfrac{\partial C}{\partial a^{-1}} & {\rm (3D)} .
\end{cases}
\end{align}
These sum rules are obeyed by our numerical results.

\subsubsection{Static bulk viscosity}

In three dimensions for negative scattering length, where no bound state exists, the zero-frequency limit of the bulk viscosity is finite. We state the exact result
\begin{align}
&\lim_{\omega\to 0} \frac{\lambda_T^3}{\hbar z^2} \zeta(\omega) \nonumber \\
&= - \frac{\sqrt{2}}{9  \pi^2} \biggl(\frac{\lambda_T}{a}\biggr)^2 \biggl[1 + e^{\lambda_T^2/2 \pi a^2} \biggl(1 + \frac{\lambda_T^2}{2 \pi a^2}\biggr) {\rm Ei}\biggl(-\frac{\lambda_T^2}{2 \pi a^2}\biggr)\biggr] , 
\end{align}
where ${\rm Ei}$ is the exponential integral. In the perturbative limit of small negative scattering length, this becomes
\begin{align}
\lim_{a \to -0 } \frac{\lambda_T^3}{\hbar z^2} \zeta &= \frac{4 \sqrt{2}}{9} \biggl(\frac{a}{\lambda_T}\biggr)^2 .
\end{align}
In the unitary limit $a \to \infty$, we obtain the result given in the Introduction, Eq.~\eqref{eq:bulkunitary}. The prefactor of $(\lambda_T/a)^2$ in Eq.~\eqref{eq:bulkunitary} is consistent with an expected scaling with $({\cal P} - \frac{2}{3} {\cal E})^2/{\cal P}^2 \sim \langle \hat{\cal O} \rangle^2/{\cal P}^2$ that vanishes in the unitary limit and measures the deviation from scale invariance~\cite{dusling13,martinez17}. Note that for the bulk viscosity in high-temperature QCD, a similar logarithmic-quadratic scaling with interaction strength is found~\cite{arnold06}. Our findings are at variance with a previous kinetic theory calculation near the unitary limit, which does not include logarithmic renormalization effects and gives a different result~\cite{dusling13}
\begin{align}
\lim_{a \to \infty} \frac{\lambda_T^3}{\hbar z^2} \zeta_{\rm kin} &= \frac{1}{24 \sqrt{2} \pi} \biggl(\frac{\lambda_T}{a}\biggr)^2 .
\end{align}
While the exact high-frequency tail and the sum rule constraints provide very compelling checks of our virial expansion and indicate the consistency of our results, a disagreement with kinetic theory is unexpected, and further work warranted.

Note that the finite result~\eqref{eq:bulkunitary} for the static bulk viscosity at unitarity is obtained only when the unitary limit of infinite scattering length ($a\to \infty$) is taken after the zero-frequency limit ($\omega \to 0$). However, these limits do not commute. By contrast, the low-frequency limit of the bulk viscosity spectral function near unitarity, i.e., taking the low-frequency limit after the unitary limit, diverges logarithmically with frequency:
\begin{align}
\lim_{\omega \to 0} \frac{\lambda_T^3}{\hbar z^2} \zeta(\omega) |_{a\to \infty} &= \dfrac{\sqrt{2} }{9 \pi^2} \biggl(\frac{\lambda_T}{a}\biggr)^2 \ln \dfrac{4 e^{-\gamma_E}}{\beta \hbar \omega} . \label{eq:bulkunitary2}
\end{align}
This result appears consistent with recent work on the nonequilibrium dynamics of a nearly unitary quantum gas~\cite{maki18,maki19a,maki19b}, which shows that the unitary conformal fixed point governs the long-time quench dynamics in these systems at times $1 \ll \ln t \ll a$. The scaling for the bulk viscosity in the time domain, $\zeta(t)\sim(\ln t)/a^2 t$  [Eq. (77) of Ref.~\cite{maki19a}], is consistent with the logarithmic low-frequency singularity~\eqref{eq:bulkunitary2}. The finite result for the static bulk viscosity quoted in the Introduction, Eq.~\eqref{eq:bulkunitary}, then corresponds to evaluating the correlation function outside of the conformal critical window.

\section{Conclusion}\label{sec:IV}

In this paper, we have presented exact nonperturbative results for the shear and bulk viscosity spectral functions calculated to quadratic order in the fugacity. The field-theoretical methods employed here allow us to make contact with many-body techniques such as the Keldysh formalism and can be compared with numerical calculations. A significant advantage of our method is that the analytic continuation from discrete Matsubara space to real frequencies is performed in closed analytical form and does not require numerical extrapolation schemes. The form of the viscosity spectral functions is linked to the two-particle scattering structure: Spectra consist of a broad continuous band superimposed with an additional bound-state contribution above the dimer-breaking threshold frequency. The results of the virial expansion are internally consistent in that they obey  universal Tan relations for the high-frequency tail and, in the case of the bulk viscosity, saturate exact sum rules.

It is interesting to note that the correspondence between the exact virial expansion in the static limit and kinetic theory is quite nontrivial. For the static shear viscosity, a memory kernel resummation of the leading low-frequency divergence in the spectral function is required to match with kinetic theory. A similar resummation scheme for the bulk viscosity is not apparent as the spectral function saturates to a finite value, albeit with an additional dimer peak where present. A direct extrapolation to the static limit for three-dimensional fermions with negative scattering length, i.e., without bound states in the two-body spectrum, gives a nonanalytic logarithmic-quadratic result near unitarity [Eq.~\eqref{eq:bulkunitary}], which differs from kinetic theory. It would be interesting to resolve this discrepancy in further work.

There are several additional applications of the methods used in this paper. For example, at the next-to-leading (third) order in the fugacity, the spectral functions include three-body effects, which provide an additional way to break scale invariance (besides an explicit finite two-body scattering length or a quantum anomaly). For the bulk viscosity of the unitary Bose gas, which is scale invariant in the two-body sector, this third-order term will be the leading nonzero order. Likewise, calculations of thermal transport coefficients will be performed in a similar way as outlined in this paper.

{\it Note added.} Recently, Refs.~\cite{enss19,nishida19} appeared, which discuss the virial expansion of the viscosity spectral functions using other methods and have overlap with this work. The overlapping results are in agreement.

\begin{acknowledgements}
I thank Sergej Moroz and Wilhelm Zwerger for discussions. I also thank Tilman Enss, Jeff Maki, Yusuke Nishida, and Fei Zhou for pointing our their work and discussions. This work was supported by Peterhouse, Cambridge.
\end{acknowledgements}

\appendix

\section{Analytic continuation}\label{app:1}

\begin{figure}[t]
\scalebox{0.5}{\includegraphics{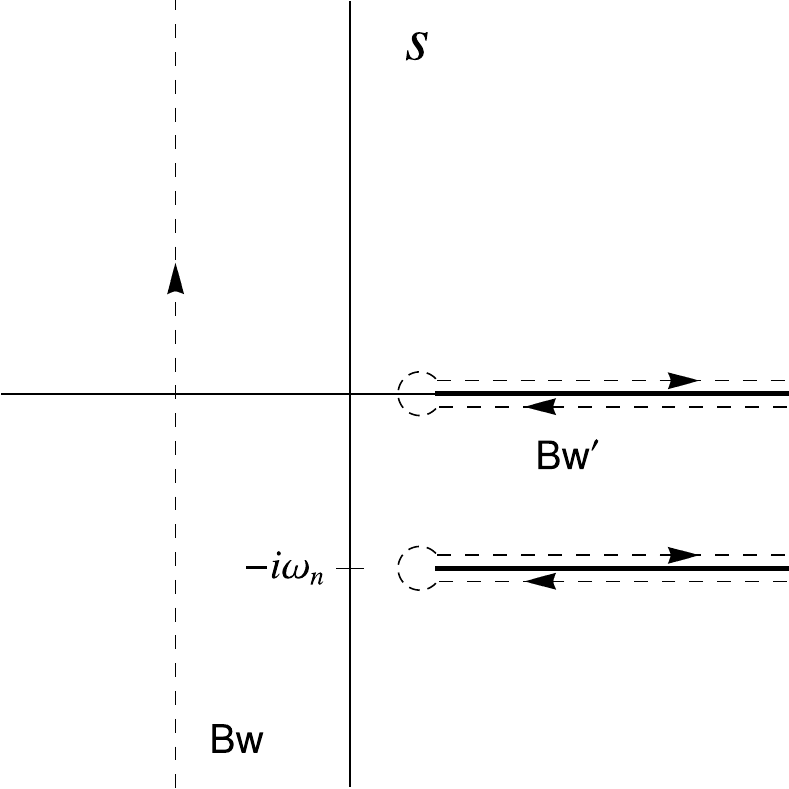}}
\caption{Contour of integration and branch cut structure used to evaluate the Feynman diagrams and to perform the analytic continuation.}
\label{fig:7}
\end{figure}

The quantum cluster expansion gives expressions for response functions in imaginary time or, after discrete Fourier transformation, in Matsubara space for frequencies $i\hbar\omega_n = 2\pi n/\beta$. Frequency-dependent retarded response functions are then obtained by analytic continuation $i\omega_n \to \omega + i0$. A considerable advantage of the cluster expansion compared to numerical methods is that this analytic continuation is performed in closed analytic form and does not require an interpolation or other numerical techniques. However, some care must be taken~\cite{mahan00}. This appendix outlines the analytic continuation for the integrals discussed in the main text.

After performing the Laplace transformation for the Feynman diagram integrals, we obtain expressions of the form
\begin{align}
\chi(i\omega_n) &= \int_{\rm Bw} \frac{ds}{2\pi i} \, e^{- \beta s} f(s) g(i\hbar\omega_n+s) , \label{eq:analytic}
\end{align}
where $f$ and $g$ are products of the single-particle propagator or the two-body $T$ matrix, which have branch cuts or poles on the real axis (we omit momentum integrals for brevity). Here Bw is the Bromwich contour, which by definition avoids every singularity of the integrand. This is shown in Fig.~\ref{fig:7}, where the black solid lines indicate the branch cuts or poles of the integrand for ${\rm Im} \, s = 0$ and ${\rm Im} (s+i\omega_n) = 0$. Deforming the contour of integration to run along the cuts ($Bw'$ in Fig.~\ref{fig:7}), we obtain
\begin{align}
\chi(i\omega_n) &= \int_{-\infty}^\infty \frac{ds}{\pi} \, e^{- \beta s} \Bigl\{ {\rm Im} [f(s+i0)] g(s+i\hbar\omega_n) \nonumber \\
&\qquad\qquad+ f(s-i\hbar\omega_n) {\rm Im} [g(s+i0)] \Bigr\} ,
\end{align}
where the integration range is extended to run along the complete real axis. We can now safely perform the analytic continuation $i\omega_n \to \omega + i0$ to obtain the retarded response function. The imaginary part reads
\begin{align}
&{\rm Im} \, \chi^R(\omega) = (1 - e^{-\beta \hbar\omega}) \nonumber \\
& \times \int_{-\infty}^\infty \frac{ds}{\pi} \, e^{- \beta s} \, {\rm Im} [f(s+i0)] {\rm Im} [g(s+\hbar\omega+i0)] . \label{eq:cont}
\end{align}
The prefactor $(1-e^{-\beta \hbar\omega})$ is expected on general grounds from the fluctuation-dissipation theorem~\cite{forster90}. For the stress correlator self-energy and Maki-Thompson diagrams, one factor $f$ or $g$ involves only simple poles from single-particle propagators, so the integration is performed using the residue theorem. The same is true for the Aslamazov-Larkin diagrams when using the Ward identity~\eqref{eq:ward}.

\section{Example calculation}\label{app:2}

\begin{figure}[t]
\scalebox{0.5}{\includegraphics{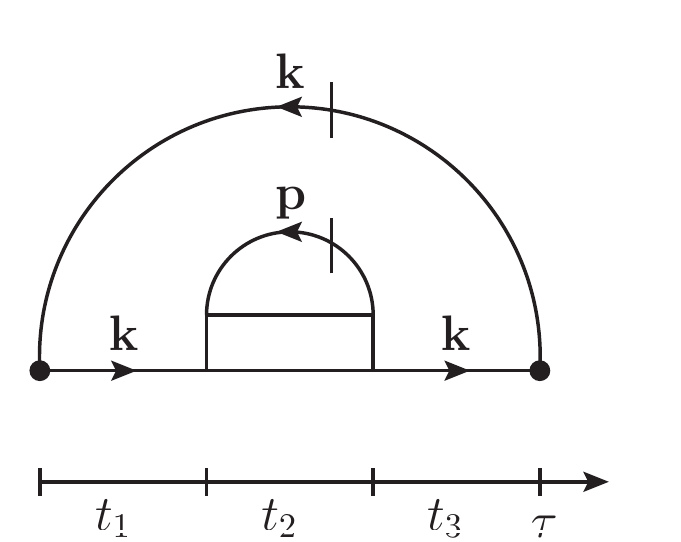}}\quad
\scalebox{0.5}{\includegraphics{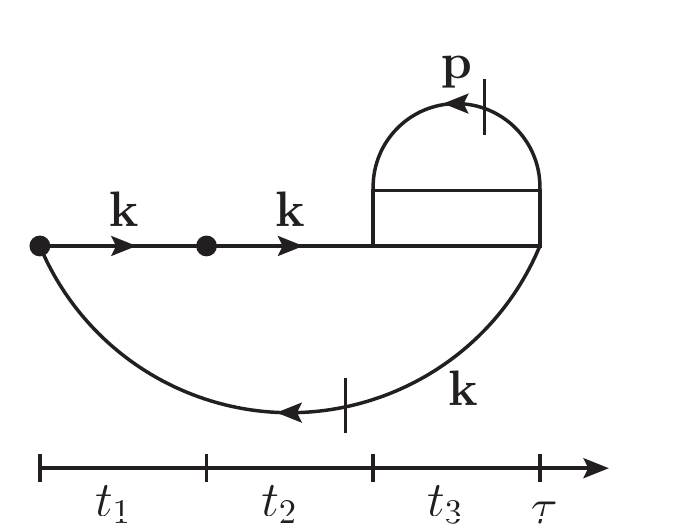}}
\caption{Self-energy contributions to the shear viscosity correlation function. Imaginary time runs from the left to the right, with insertions of the off-diagonal stress tensor at imaginary times $0$ and $\tau$ (black dots).}
\label{fig:8}
\end{figure}

To illustrate the diagrammatic framework discussed in the main text, this appendix presents the calculation of the self-energy contribution to the shear viscosity spectral function in detail. The corresponding Feynman diagrams are shown in Fig.~\ref{fig:3}(b) and also in Fig.~\ref{fig:8} including the imaginary-time axis. The calculation of other diagrams proceeds in a similar way. 

The Feynman diagrams discussed in this paper lie on an imaginary-time axis, with imaginary time running from the left-hand side to the right-hand side. We partition the interval $[0,\tau]$ in three parts with interval lengths $t_1$,$t_2,$ and $t_3$ as shown in Fig.~\ref{fig:8}, where $\tau = t_1 + t_2 + t_3$. Both diagrams in Fig.~\ref{fig:8} contain two retarded propagators, both with momentum ${\bf k}$. Using the Feynman rules in Eq.~\eqref{eq:feynmanprop}, they contribute factors of $[- \Theta(t_1) \, e^{- t_1 \varepsilon_{\bf k}}]$ and $[- \Theta(t_3) \, e^{- t_3 \varepsilon_{\bf k}}]$ for the first diagram and $[- \Theta(t_1) \, e^{- t_1 \varepsilon_{\bf k}}]$ and $[- \Theta(t_2) \, e^{- t_2 \varepsilon_{\bf k}}]$ for the second diagram (we neglect an additional factor $e^{\tau \mu}$, the argument of which will add to zero as all fermion lines are closed). The diagrams contain two-body scattering matrices, which sum all repeated scattering events shown in Fig.~\ref{fig:2b} and which contribute terms of $[\Theta(t_2) T_2(t_2, {\bf k} + {\bf p})]$ and $[\Theta(t_3) T_2(t_3, {\bf k} + {\bf p})]$, respectively. The overall leading order in the fugacity of these diagrams is determined by the backward propagating lines, which are ${\cal O}(z)$, such that the total diagram contributes at order ${\cal O}(z^2)$ to the response function. The advanced Green's functions contribute factors $[z e^{- (\beta - t_1 - t_2 - t_3) \varepsilon_{\bf k}}]$ and $[z e^{- (\beta - t_2) \varepsilon_{\bf p}}]$ or $[e^{- (\beta - t_2) \varepsilon_{\bf p}}]$, respectively. Having evaluated these factors, we integrate over the undetermined loop momenta ${\bf k}$ and ${\bf p}$ and take the discrete Fourier transform to obtain the response in Matsubara space.

Collecting all terms, the first Feynman diagram evaluates to
\begin{align}
&\chi_{xy,xy}^{\rm SE,a}(i \omega_n) = 2 z^2 \int d(t_1,t_2,t_3) e^{i \hbar \omega_n (t_1+t_2+t_3)} \nonumber \\
&\times \int \frac{d^dp}{(2\pi)^d} \frac{d^dk}{(2\pi)^d} [\Theta(t_1) e^{- t_1 \varepsilon_{{\bf k}}}] [\Theta(t_2) T_2(t_2, {\bf k} + {\bf p})] \nonumber \\
& \times  [\Theta(t_3) e^{- t_3 \varepsilon_{{\bf k}}}] [e^{- (\beta - t_1 - t_2 - t_3) \varepsilon_{\bf k}}] \, [e^{- (\beta - t_2) \varepsilon_{\bf p}}] \nonumber \\
&\times \, \Theta(\beta - t_1 - t_2 - t_3)  T^0_{xy}({\bf k}) T^0_{xy}({\bf k}) ,
\end{align}
where the prefactor of $2$ accounts for a spin summation. Rearranging the factors gives
\begin{align}
&\chi_{xy,xy}^{\rm SE,a}(i \omega_n) = z^2 \int \frac{d^3p}{(2\pi)^3} \int \frac{d^3k}{(2\pi)^3} \, T^0_{ij}({\bf k}) T^0_{kl}({\bf k}) \nonumber \\
&\quad \times   \int d(t_1,t_2,t_3) \, [\Theta(t_1) e^{i\hbar \omega_n t_1} e^{- t_1 (\varepsilon_{\bf k} + \varepsilon_{\bf p})}] \, \nonumber \\
&\quad \times  [\Theta(t_2) e^{i\hbar \omega_n t_2} T_2(t_2, {\bf k} + {\bf p})]  [\Theta(t_3) e^{i\hbar \omega_n t_3} e^{- t_3 (\varepsilon_{\bf k} + \varepsilon_{\bf p})}] \nonumber \\
&\quad \times   [\Theta(\beta - t_1 - t_2 - t_3) e^{- (\beta - t_1 - t_2 - t_3) (\varepsilon_{\bf k} + \varepsilon_{\bf p})}] .
\end{align}
We now apply the convolution theorem~\eqref{eq:convolution}, which gives the expression~\eqref{eq:se_a}. The second line of Eq.~\eqref{eq:se_a} takes the form~\eqref{eq:analytic} with
\begin{align}
f(s) &= \frac{1}{s - \varepsilon_{{\bf k}} - \varepsilon_{{\bf p}}} , \\
g(s) &= \frac{T_2(s, {\bf k} + {\bf p})}{(s - \varepsilon_{{\bf k}} - \varepsilon_{{\bf p}})^2} .
\end{align}
Performing the analytic continuation as outlined in Appendix~\ref{app:1}, we end up with an expression of the form~\eqref{eq:cont}. As mentioned above, this expression is further simplified using ${\rm Im} [f(s+i0)] = - \pi \delta(s - \varepsilon_{{\bf k}} - \varepsilon_{{\bf p}})$, which removes the $s$ integration. This then yields the expression~\eqref{eq:res_chiSEa}.

Repeating the same calculation for the second diagram in Fig.~\ref{fig:8}, the Bromwich integral takes a similar form as before with the functions $f$ and $g$ interchanged:
\begin{align}
f(s) &= \frac{T_2(s, {\bf k} + {\bf p})}{(s - \varepsilon_{{\bf k}} - \varepsilon_{{\bf p}})^2} \\
g(s) &= \frac{1}{s - \varepsilon_{{\bf k}} - \varepsilon_{{\bf p}}} .
\end{align}
Evaluating the residue of $g(s)$ then gives the expression~\eqref{eq:res_chiSEb}.

\bibliography{bib}

\end{document}